\documentclass[useAMS,usenatbib]{mn2e}
\usepackage{epsfig}
\usepackage{epstopdf}
\usepackage{graphicx}
\usepackage{amsmath}
\usepackage{amssymb}
\usepackage{natbib}
\usepackage[usenames]{color}
\newcommand{\kms}{\ensuremath{{\rm km\,s}^{-1}}}

\newcommand{\K}{\ensuremath{{\rm K}}}
\newcommand{\msun}{\ensuremath{{\rm M_{\odot}}}}
\newcommand{\hp}{\ensuremath{{h_{\rm p}}}}
\newcommand{\kb}{\ensuremath{{k_{\rm b}}}}
\newcommand{\mH}{\ensuremath{{m_{\rm p}}}}

\newcommand{\Sgas}{\Sigma_{\rm g}}
\newcommand{\dSstar}{\dot{\Sigma}_{\star}}
\newcommand{\dmdisk}{\dot{M}_{\rm d}}
\newcommand{\bdmdisk}{\dot{M}_{\rm d, ave}}
\newcommand{\rdisk}{R_{\rm d}}
\newcommand{\mcl}{M_{\rm cl}}
\newcommand{\mgmc}{M_{\rm H_{2}}}
\newcommand{\rcl}{R_{\rm cl}}
\newcommand{\rco}{R_{\rm CO}}
\newcommand{\rgmc}{R_{\rm H_{2}}}
\newcommand{\fgmc}{f_{\rm H_{2}}}
\newcommand{\fco}{f_{\rm CO}}
\newcommand{\Scl}{\Sigma_{\rm cl}}
\newcommand{\ncl}{N_{\rm cl}}
\newcommand{\mach}{\mathcal{M}}
\newcommand{\pfunc}{\mathcal{Z}}
\newcommand{\sigcl}{\sigma_{\rm cl}}

\newcommand{\cs}{c_{\rm s}}
\newcommand{\tgas}{T_{\rm gas}}
\newcommand{\texc}{T_{\rm exc}}
\newcommand{\tcmb}{T_{\rm CMB}}
\newcommand{\nuco}{\nu_{\rm CO}}
\newcommand{\ncrit}{n_{\rm crit}}
\newcommand{\nmed}{n_{\rm med}}
\newcommand{\nave}{\bar{n}_{\rm cl}}
\newcommand{\nco}{\bar{n}_{\rm CO}}
\newcommand{\xco}{X_{\rm CO}}
\newcommand{\Fco}{F_{\rm CO}}
\newcommand{\Fcmb}{F_{\rm CMB}}
\newcommand{\muv}{M_{\rm UV}}
\newcommand{\mhalo}{M_{\rm h}}

\begin{document}

\title[GMCs and CO at $z \gtrsim 6$]{Molecular Cloud Properties and CO Line Emission in $z \gtrsim 6$ Galaxies}

\author[Mu{\~n}oz and Furlanetto]
{
Joseph A.\ Mu{\~n}oz\thanks{E-mail: jamunoz@astro.ucla.edu} 
and 
Steven R. Furlanetto
\\University of California Los Angeles, Department of Physics and Astronomy; Los Angeles, CA 90095, USA
}

\maketitle

\begin{abstract}
We explore molecular cloud properties and the physics of CO transition lines in $z\gtrsim6$ Lyman-break galaxies and predict their CO fluxes using an analytic formalism built from global models of star formation in high-redshift galaxies that minimizes our reliance on local observations.  Our model includes a new approach to calculating the molecular gas fraction that takes the total gas density, the star formation rate, and the star formation efficiency in clouds as the principal inputs.  This method agrees with chemical equilibrium calculations of the molecular fraction based on local chemistry if galaxies at $z\gtrsim6$ have metallicities of order a few percent of solar.  Such low metallicities in turn imply that much of the carbon in these systems exists in ionized form rather than as CO. Moreover, we find that the higher-order CO transitions observable at high redshift with ALMA will typically be sub-thermally populated but that the details depend sensitively on the presence of turbulent clumps within molecular clouds.  Ultimately, we expect current facilities will only be able to observe the CO signal from reionization epoch galaxies with great difficulty.  We estimate that at least $\sim100$ hours of integration time with ALMA will be required to detect the CO(6--5) transition in $z=6$ systems with rest-frame UV magnitudes of -20.
\end{abstract}


%-------------------------------------------------------------------------------------------------------------
%             Introduction
%-------------------------------------------------------------------------------------------------------------
\section{Introduction}

The WFC3 camera aboard the {\it{Hubble Space Telescope}} has discovered large populations of galaxies out to $z\sim9$ \citep[e.g.,][]{Bouwens06, Bunker10, McLure10, Finkelstein10, Bouwens11a, Bouwens11b, Ellis13} via the Lyman-break technique and opened investigation into galaxy formation during the first $\sim500\,{\rm Myrs}$ of cosmic time.  Observations in the rest-frame optical and UV have been important in estimating star formation rates and stellar masses for these Lyman-break galaxies (LBGs) in addition to the census statistics of abundance and clustering.  According to such measurements, typical galaxies during the epoch of reionization at $z \gtrsim 6$ are much smaller and denser than those at low redshift \citep[e.g.,][]{Oesch10b, Ono13} and would be considered dwarfs based on their stellar masses \citep[e.g.,][]{Labbe10b}.  However, these observations tell us little about the thermal and chemical state of high-redshift interstellar media (ISM).

On the other hand, observations of molecular transition lines expected from the the Jansky Very Large Array (JVLA) and the Atacama Large Millimeter Array (ALMA), particularly those of CO and CII, will probe internal galactic physics and dynamics \citep[e.g.][]{Carilli08, WC08, CW13}.  Averaged over large scales, intensity maps of these lines can also validate and enhance 21 cm studies of reionization \citep[e.g.][]{Lidz11, Carilli11, Gong11, Gong12}.

In this {\it{Paper}}, we probe the relationship between the conditions expected in the high-redshift ISM and the physics of CO transitions using the model for high-redshift galactic disks we developed in \citet{MF12}.  Additionally, we aim to make predictions for the CO signal from these sources to facilitate current and future observations.  We distinguish the systems under discussion in this paper from those quasars and sub-millimeter galaxies in which CO lines have already been observed \citep[see][for a recent review]{CW13}.  This latter classes of objects are much rarer than the galaxies of interest here, and we plan to address them separately in forthcoming work.

Recent attempts at predicting the CO emission signal from high-redshift galaxies \citep[e.g.,][]{Obreschkow09b, Lidz11} have relied heavily on empirical relations calibrated from local observations.  For example, \citet{Obreschkow09b} used a semi-analytic model of galaxy formation \citep{Croton06, dLB07} to consider galaxy sizes and masses appropriate to high-redshift while including the effect of the cosmic microwave background (CMB)---whose temperature is nearly $20\,{\rm K}$ by $z=6$---as both a source of heating and as an observational background.  However, the galaxy formation prescriptions of their semi-analytic model, the empirical pressure-relation that sets their molecular fractions, their stellar and AGN heating mechanisms, the optical depths of their excitation states, their assumed dependence on metallicity, and the normalization factor of their CO luminosities were all calibrated from low-redshift measurements.  Such studies have primarily lacked a detailed model for giant molecular clouds (GMCs) in the ISM of high-redshift galaxies that describes how molecular fractions, GMC sizes and densities, gas temperatures, and molecular level populations depend on halo mass, redshift, and metallicity.

In principle, these questions can be addressed by numerical simulations.  \citet{Narayanan12} has conducted simulations of galaxies in isolation out to $z\sim2$ with a modified version of the smoothed-particle hydrodynamic code GADGET2 coupled to the radiative transfer code SUNRISE to develop a general model for $\xco$---the ratio of molecular gas density to CO emission temperature---including physically reasonable prescriptions for the effect of metallicity, molecular fraction, and gas temperature.  However, their work lacks some of the important elements necessary for the calculation at $z\gtrsim6$, namely, the ability to extrapolate to the higher CO excitation states observable with ALMA at high-redshift, the consideration of an (often dominant) CMB background, and a cosmological context for their isolated galaxies.  Moreover, it is unclear whether the sub-grid prescriptions for thermal supernova feedback in the GADGET code are appropriate in the early universe as they are also tuned to reproduce observations at $z=0$.

Ultimately, modern simulations can still only probe a narrow dynamic range of size scales.  Simultaneously capturing the full cosmological context of halo growth and the physics of photo-dissociative regions (PDRs) within GMCs is typically outside their grasp.  Our approach is to use a sophisticated analytic framework that pieces together physically-reasonable models at all scales.  While some of our assumptions will be quite idealized, they capture the necessary physics to achieve our goals of understanding the physics of CO transitions and predicting their flux in a high-redshift context while minimizing our reliance on low-redshift observations.  

In building our formalism, we will work within a paradigm where star formation is a function of gas supply and stellar feedback \citep[e.g.,][]{HQ11, Hopkins11, Dave12}.  We begin with a relation between star formation, cold-flow accretion, and momentum-driven feedback \citep{Dave12} that is known to accurately describe the star formation rates and UV luminosities of $z\gtrsim6$ LBGs, their dependence on halo mass, and their evolution with redshift \citep{Munoz12}.  We incorporate these inflow and outflow rates into a model for radiation-pressure--supported galactic disks originally developed to describe $z\sim2$ starbursts and AGN \citep{Thompson05} but improved and extended to apply to higher-redshift systems \citep{MF12}.  This disk model provides radial distributions of star formation rate, gas density, and effective temperature for assumptions about the angular momentum transport mechanism within the disk (\S\ref{sec:disk}).  We use these quantities to calculate the masses, sizes, and number of GMCs as a function of disk radius (\S\ref{sec:H2:gmcs}) and propose a new method for determining molecular fractions such that the star formation law is obeyed at each radius given the star formation rates and gas densities prescribed by our framework (\S\ref{sec:H2:fH2}).  On the smallest scales, we consider PDRs within GMCs, the geometric extent of CO-enriched gas (\S\ref{sec:H2:fco}), the gas temperature within clouds (\S\ref{sec:temp}), and the state of the rotational excitation levels (\S\ref{sec:excitation}).  In \S\ref{sec:Lco}, we compute the total CO line luminosity from the properties calculated in the previous sections by counting the number of clouds as a function of galactic radius.  We compare our predictions for $\xco$ to the \citet{Narayanan12} model in \S\ref{sec:Xco} and, in \S\ref{sec:results}, make predictions for the CO luminosity function (\S\ref{sec:results:COLF}) and for followup observations of UV-selected targets (\S\ref{sec:results:CO-UV}).  In \S\ref{sec:comp}, we compare our results with those from previous studies, particularly from the works of \citet{Obreschkow09b} and \citet{Lagos12}.  Finally, we summarize and conclude in \S\ref{sec:conclusions}.

%-------------------------------------------------------------------------------------------------------------
%            Disk Model
%-------------------------------------------------------------------------------------------------------------
\section{Galactic Disks}\label{sec:disk}

In \citet{MF12}, we developed a one-zone model for the ISM of galaxies at $z\gtrsim6$.  Here, we briefly outline the model far enough to calculate the quantities relevant for the present work, but we refer the reader to our previous paper for further details.  Our formalism is based on work by \citet{Thompson05} but improved and tailored specifically to describe $z\gtrsim6$ systems. 

The model assumes a rotationally-supported galactic disk in which marginal Toomre-instability (i.e., $Q=1$) and vertical hydrostatic equilibrium are maintained by radiation pressure from stars and mechanical pressure from supernovae.  Embedded in an isothermal halo of mass $\mhalo$ at redshift $z$ whose gravity dominates the rotation, the disk has a gas density as a function of radius given by
\begin{equation}\label{eq:rho}
\rho=\frac{B^{\frac{1}{2}}\,\Omega^2}{\sqrt{2}\,\pi\,G\,Q},
\end{equation}
where $\Omega=\sqrt{2}\,\sigma/r$ and $\sigma$ is the halo velocity dispersion.  $B-1\approx8.9$ is the ratio of the vertical self-gravity of the disk to that of the halo.  The total vertical pressure is
\begin{equation}\label{eq:pressure}
P_{\rm tot}=B\,\rho\,h^2\,\Omega^2,
\end{equation}
where $h$ is the vertical scale height.  The disk sound speed is, thus, given by $\cs=B^{1/2}\,h\,\Omega$.  The thickness of the disk as a function of radius is set by the amount of gas flowing through it toward the center:
\begin{equation}\label{eq:h}
h=\frac{\dot{M}}{4\,\pi\,r\,\rho\,v_{\rm in}},
\end{equation}
where $v_{\rm in}$ is the inflow velocity and $\dot{M}$ is the mass inflow rate.  Additionally, we take the maximum disk radius, $\rdisk$, to be a fixed fraction $\lambda/\sqrt{2}$ of the halo virial radius with $\lambda=0.05$ \citep{MMW98, WL11a}.

The physical properties of the disk are thus determined by the mass and redshift of the host halo and the inflow of gas through the disk as a function of radius.  We set the boundary condition on the latter so that the infall rate onto the edge of the disk is given by the cold flow accretion rate, $\dmdisk$, the average of which as a function of halo mass and redshift is \citep{McBride09}
\begin{equation}\label{eq:acc_cf} 
\bdmdisk \approx 3\,{\rm \msun/yr}\,\left(\frac{\mhalo}{10^{10}\,\msun}\right)^{1.127}\,\left(\frac{1+z}{7}\right)^{2.5}\,\left(\frac{f_{\rm b}}{0.16}\right),
\end{equation}
where $f_{\rm b}$ is the cosmic baryon fraction.  This choice provides a cosmological context for our galaxy models fed by the growth of large-scale structure.  At each radius from the edge, star formation and outflows deplete the amount of inflowing gas by an amount $2\,\pi\,r\,\dSstar\,(1+\eta_{\rm wind})\,dr$, where $\dSstar$ is the star formation rate surface density and $\eta_{\rm wind}$ is the wind mass-loading factor.  We found in \citet{MF12} that momentum transfer from supernovae and ionizing photons contribute nearly equally to hydrostatic equilibrium.  Where pressure from these stellar sources dominates thermal gas pressure, the star formation rate surface density is
\begin{equation}\label{eq:SFR}
\dSstar=\frac{B\,\rho\,h^2\,\Omega^2}{2\,\epsilon\,c},
\end{equation}
where the efficiency with which star formation converts baryons into radiation, $\epsilon \approx 10^{-3}$, is only weakly dependent on IMF.  Moreover, we parameterize $\eta_{\rm wind}$ as owing to momentum-driven winds,
\begin{equation}\label{eq:eta_wind}
\eta_{\rm wind}=\eta_{0}\,\frac{100\,{\rm km/s}}{\sigma}\,\frac{\epsilon}{10^{-3}},
\end{equation}
and set $\eta_{0}=4$ to match model fits of the high-z UV luminosity function \citep{Munoz12} and in agreement with numerical simulations \citep{OD08}.  We further assume that star formation heats the gas to a temperature, $T$, given by 
\begin{equation}\label{eq:temp}
\sigma_{\rm SB}\,T^4 = \frac{1}{2}\,\dSstar\,\epsilon\,c^2
\end{equation}
such that, in the inner regions of the disk---but still well outside the accretion disk of a central black hole---thermal gas pressure,
\begin{equation}\label{eq:pgas}
P_{\rm gas}=\frac{\rho\,\kb\,T}{\mH},
\end{equation}
may contribute to the maintenance of vertical hydrostatic equilibrium.

Finally, to characterize the inflow velocity, $v_{\rm in}$, as a function of disk radius we consider two different phenomenological models of angular momentum transport, both of which appeal to global gravitational torques rather than local processes.  The first is a linear spiral wave (LSW) mode, where $v_{\rm in}$ is allowed reach a fixed fraction of the local sound speed \citep[e.g.,][]{Kalnajs71, LBK72, Goodman03}:
\begin{equation}\label{eq:vin_m}
v_{\rm in}=m\,\cs.
\end{equation}
By contrast, we also consider a scenario in which non-linear inflow due to shocks and orbit-crossings gives a velocity that is a constant fraction, $\beta$, of the local circular velocity \citep[e.g.,][]{PP77, BT87, HQ11}:
\begin{equation}\label{eq:vin_s}
v_{\rm in}=\beta\,r\,\Omega.
\end{equation}
We showed in \citet{MF12} that inflow driven by a local $\alpha$-viscosity \citep{SS73} cannot supply enough gas to the center of the disk to grow a black hole.  Therefore, we ignore such a model in the present work.

\begin{figure}
\begin{center}
\includegraphics[width=\columnwidth,trim=0 0 210 0,clip]{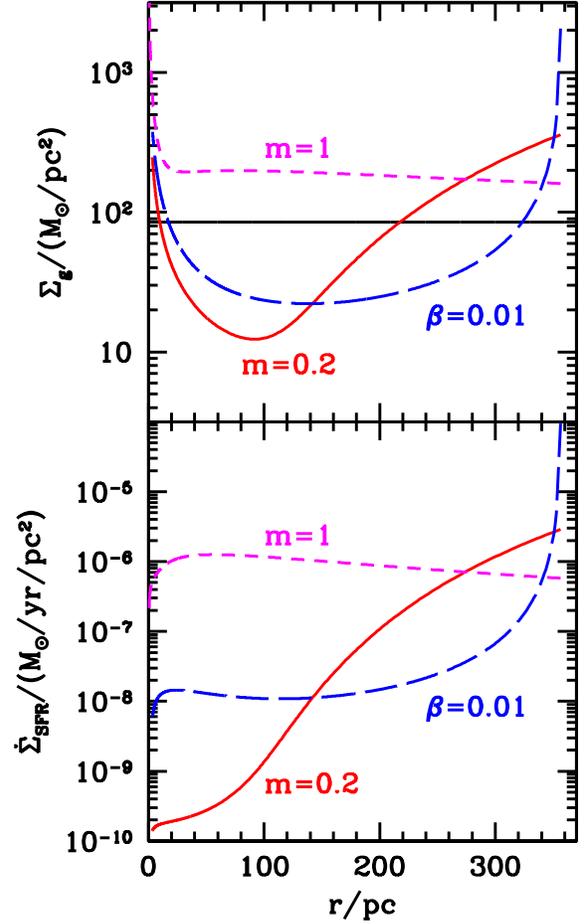}
\caption{\label{fig:sigma_sfr}
Disk properties from the model described in \S\ref{sec:disk} for a halo mass of $10^{10}\,\msun$ at $z=6$.  The top panel shows the gas surface density as a function of radius for three different angular momentum transport models: LSW models with $m=0.2$ (solid, red) and 1 (short-dashed, magenta) and nonlinear infall with $\beta=0.01$ (long-dashed, blue).  The horizontal line marks $85\,{\rm M_{\odot}/pc^2}$, the minimum surface density of GMCs.  The lower panel plots the star formation surface density for the same fiducial models.
}
\end{center}
\end{figure}

Combining equations \ref{eq:rho}-\ref{eq:vin_s}, the model outlined in this section gives $\dSstar$, $\rho_{\rm g}$, $h$, and $T$ as functions of radius for a specified halo mass and redshift.  Figure \ref{fig:sigma_sfr} shows the gas and star formation rate surface densities as functions of galactic disk radius in $10^{10}\,\msun$ halos at $z=6$ for three fiducial angular momentum transport models: LSW models with $m=0.2$ and 1 and a shocked infall model with $\beta=0.01$.  Results for a ``high" inflow shocked model with $\beta=0.1$ are similar to those for the LSW model with $m=1$.  Note that, since each of these models has the same halo mass and redshift, the total gas masses and star formation rates are roughly the same in each case.  Models in which gas is transported to the center of the disk more quickly produce radial distributions of galaxy properties that are more uniform throughout the disk because the gas is not as quickly depleted by star formation and winds.  This type of steady-state, disk model is likely appropriate for LBGs during the epoch of reionization since mergers are not expected to be a dominant influence \citep[e.g.,][]{Pawlik11, Dave12}.  In the following sections, we will use these disk properties to characterize GMCs in high-redshift galaxies and calculate their CO line-luminosities.

%-------------------------------------------------------------------------------------------------------------
%             The Molecular Gas
%-------------------------------------------------------------------------------------------------------------
\section{The Molecular Gas}\label{sec:H2}

We now consider the properties of the molecular gas in the disks described in \S\ref{sec:disk}.  While we use the results from the previous section---which assumed strict cylindrical symmetry---we focus our attention on smaller scales where the gas is clumpy and arranged into clouds.  We adopt the \citet{Wolfire10} picture of multilayered photo-dissociative regions (PDRs) within clouds (see the schematic diagram in their Fig. 1) and calculate GMC properties (\S\ref{sec:H2:gmcs}), the molecular fraction (\S\ref{sec:H2:fH2}), and the dark gas fraction (\S\ref{sec:H2:fco}) as functions of radius through the disks of high-redshift galaxies.  As is typical in the recent literature of GMCs in high-redshift galaxies \citep[e.g.,][]{Narayanan12}, we assume that all clouds in regions where $\Sgas < 85\,{\rm \msun\,pc^{-2}}$ have constant and uniform surface densities of $\Scl=85\,{\rm \msun\,pc^{-2}}$.  Otherwise, $\Scl=\Sgas$.  Below this value, internal self-regulation of GMCs dominated by feedback from HII regions reaches equilibrium at about $100\,{\rm \msun\,pc^{-2}}$ \citep{Krumholz06}.  However, at higher values of $\Sgas$, the external pressure on GMCs grows larger than the internal pressure, and the column density must increase to compensate \citep{Krumholz09b}.  This threshold is plotted against $\Sgas$ for a $10^{10}\,\msun$ halo in the top panel of Figure \ref{fig:sigma_sfr}.  Note that most of the gas mass and star formation occur in the outer portions of the disks where the GMCs will typically have higher surface densities than those found locally.

%-------------------------------------------------------------------------------------------------------------
%             Properties of Giant Molecular Clouds
%-------------------------------------------------------------------------------------------------------------
\subsection{Properties of Giant Molecular Clouds}\label{sec:H2:gmcs}

\begin{figure*}
\begin{center}
\includegraphics[width=6in]{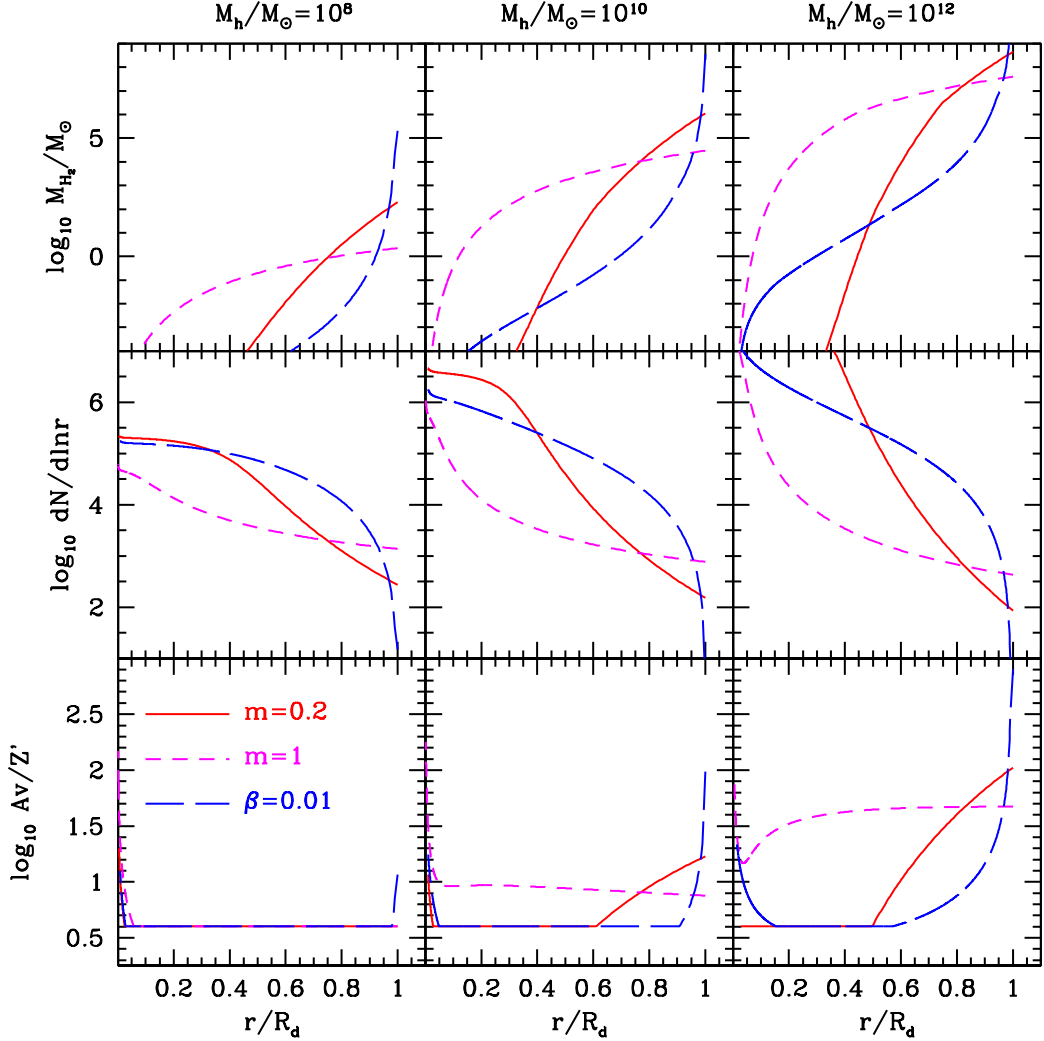}
\caption{\label{fig:gmc} 
GMC properties for our fiducial angular momentum transport models at $z=6$.  The top, middle, and bottom panels show the GMC mass, the abundance of clouds, and the visual extinction through a cloud, respectively, as functions of galactic radius, while left, center, and right columns assume host halos of $10^{8}$, $10^{10}$, and $10^{12}\,{\msun}$.  As in Fig. \ref{fig:sigma_sfr}, solid (red), short-dashed (magenta), and long-dashed (blue) curves correspond to LSW models with $m=0.2$ and 1 and a nonlinear infall model with $\beta=0.01$, respectively.
}
\end{center}
\end{figure*}

In this section, we calculate the mass, number, and visual extinction of GMCs in $z\gtrsim6$ galaxies.  We assume that all gas in the disk fragments into atomic-molecular complexes with masses of order the local Jeans mass:
\begin{equation}\label{eq:Mcl}
\mcl=M_{\rm Jeans}=\frac{\pi^4\,G^2\,\Sgas^3\,Q^4}{4\,B\,\Omega^4},
\end{equation}
where $B\approx9.9$ is defined in \S\ref{sec:disk}.  Note that this mass includes both molecular and atomic gas as the fragmentation process does not distinguish between the two.  The typical molecular mass within one of these clouds is given by $\mgmc=\fgmc\,\mcl$, where $\fgmc$ is the molecular fraction of the disk.  This framework implicitly assumes that all gas fragments in this way and, in particular, that there is no smooth molecular component.  We present our new method for determining $\fgmc$ in \S\ref{sec:H2:fH2}, but for the purpose of calculating cloud properties in this section we give the result as
\begin{equation}\label{eq:fH2}
\fgmc ={\rm min}\begin{cases} \left[\left(\frac{0.68\,{\rm Gyr}}{\Sgas/\dSstar}\right)\,M_{\rm cl, 6}^{0.33}\,\Sigma_{\rm cl, 85}^{-0.67}\,\dot{\Sigma}_{\rm \star, 1}^{-0.04}\right]^{1.33}\\1\end{cases},
\end{equation}
where $M_{\rm cl, 6}=\mcl/10^6\,\msun$, $\Sigma_{\rm cl, 85}=\Scl/(85\,{\rm \msun\,pc^{-2}})$, and $\dot{\Sigma}_{\rm \star, 1}=\dSstar/({\rm \msun\,yr^{-1}\,kpc^{-2}})$.  The top panels of Figure \ref{fig:gmc} show the variation in cloud mass with fractional galactocentric radius within $z=6$ galaxies for our fiducial angular momentum transport models over a wide range of halo masses.  Generally, the lower gas surface densities and increased angular velocities in the central portions of a disk cause the Jeans mass---and so the GMC mass---to plummet from the outer disk edge to the disk center.  The cloud mass eventually drops bellow a solar mass so that stars themselves should have difficulty fragmenting out of the disk.  While this may be unphysical and demonstrate a limitation of our model, hydrostatic equilibrium in these regions is mostly maintained by thermal pressure rather than by stellar feedback.  Thus, the small cloud masses do not undermine our calculation of disk properties there.

Atomic-molecular complexes have a typical radius of $\rcl \sim \sqrt{\mcl/(\pi\,\Scl)}$ such that, when $\Sgas > 85\,{\rm \msun\,pc^{-2}}$, $\rcl \sim h$.  Following \citet{Wolfire10}, the average number density within a cloud is $\nave=3\,\mcl/(4\,\pi\,\rcl^3\,\mu_{\rm H})=3\,\ncl/(4\,\rcl)$, where $\ncl$ is the column density of the cloud and $\mu_{\rm H}=2.34\times10^{-24}\,{\rm grams}$ is the proton mass.  The cloud surface density, $\Scl$, is assumed to be constant with radius, $r_{\rm cl}$, within the cloud, and thus the average density scales as $\bar{n} \propto r_{\rm cl}^{-1}$.  This means that the mass enclosed within $r_{\rm cl}$ is $M(r_{\rm cl}) \propto r_{\rm cl}^2$.  Thus, if all clouds have the same surface density, then the total surface area of all clouds is proportional to the total mass in clouds.  In the optically thick limit of CO emission, this results in a luminosity that is proportional to gas mass.  As a further consequence, the typical GMC radius, i.e., the radius of the cloud containing molecular gas, is $\rgmc=\fgmc^{1/2}\,\rcl$.

The distribution of clouds throughout the disk, $dN/d{\ln}r$, is given by the ratio of the total molecular mass at a given galactocentric radius to the GMC mass:
\begin{equation}\label{eq:N}
\frac{dN}{d{\ln}r}(r)=\frac{2\,\pi\,r^2\,\fgmc\,\Sgas}{\mgmc}=\frac{2\,\pi\,r^2\,\Sgas}{\mcl}.
\end{equation}
Substituting for $\mcl$ from equation \ref{eq:Mcl} yields
\begin{equation}\label{eq:Nreduced}
\frac{dN}{d{\ln}r}(r)=\frac{8\,B\,\Omega^4\,r^2}{\pi^3\,G^2\,\Sgas^2\,Q^4}.
\end{equation}
As shown in the central row of panels in Figure \ref{fig:gmc}, the number of GMCs increases strongly toward the center of the disk due to the rapid drop in cloud mass.  

The average visual extinction coefficient within the GMC, $A_{V}$, will be important in \S\ref{sec:H2:fco} when considering how deeply dissociating radiation penetrates into the cloud.  Following \citet{Wolfire10}, we assume, for simplicity, that the dust-to-gas ratio is constant within the cloud and that the dust opacity scales with metallicity.  Normalizing the extinction to be unity for a column density of $1.9\times10^{21}\,{Z'}^{-1}\,{\rm cm^{-2}}$ gives 
\begin{equation}
A_{V}=\frac{\ncl\,Z'}{1.9\times10^{21}\,{\rm cm^{-2}}},
\end{equation}
where $Z'=Z/Z_{\odot}$ is the metallicity in solar units.  As shown in Figure \ref{fig:gmc}, where $\Sgas > 85\,{\rm \msun\,pc^{-2}}$, $A_{V}\propto\Sgas$.  Otherwise, $A_{V}$ is constant.

%-------------------------------------------------------------------------------------------------------------
%             The Molecular Fraction
%-------------------------------------------------------------------------------------------------------------
\subsection{The Molecular Fraction}\label{sec:H2:fH2}

In this section, we consider a new approach for calculating the molecular gas fraction in the disk, $\fgmc=\mgmc/\mcl$, that attempts to reconcile two opposing views of star formation.  On the one hand, \citet{Krumholz09b} express star formation as a purely local process depending critically on the amount of molecular hydrogen present at a given location:
\begin{equation}\label{eq:SFlaw}
\dSstar=\Sgas\,\fgmc\, \frac{{\rm SFR_{\rm ff}}}{t_{\rm ff}},
\end{equation}
where ${\rm SFR_{\rm ff}}/t_{\rm ff}$ is the star formation efficiency per free-fall time divided by the free-fall time.  After modeling $\fgmc$ and the efficiency as the fundamental physical inputs, they calculate $\dSstar$ as the derived quantity.  In contrast, \citet{Hopkins12a} find that galactic star formation depends on a global competition between the infalling supply of cold gas and feedback and negligibly on local molecular gas chemistry \citep[see also][]{Dave12}; this latter perspective was implicitly assumed in the model presented in \S\ref{sec:disk}.  In our calculation of the molecular fraction, we take $\dSstar$ as the primary input based on the results of our disk model.  Then, calculating ${\rm SFR_{\rm ff}}/t_{\rm ff}$ from the GMC properties determined in \S\ref{sec:H2:gmcs}, we solve for the value of $\fgmc$ required to produce this star formation from equation \ref{eq:SFlaw}.

\begin{figure*}
\begin{center}
\includegraphics[width=6.0in]{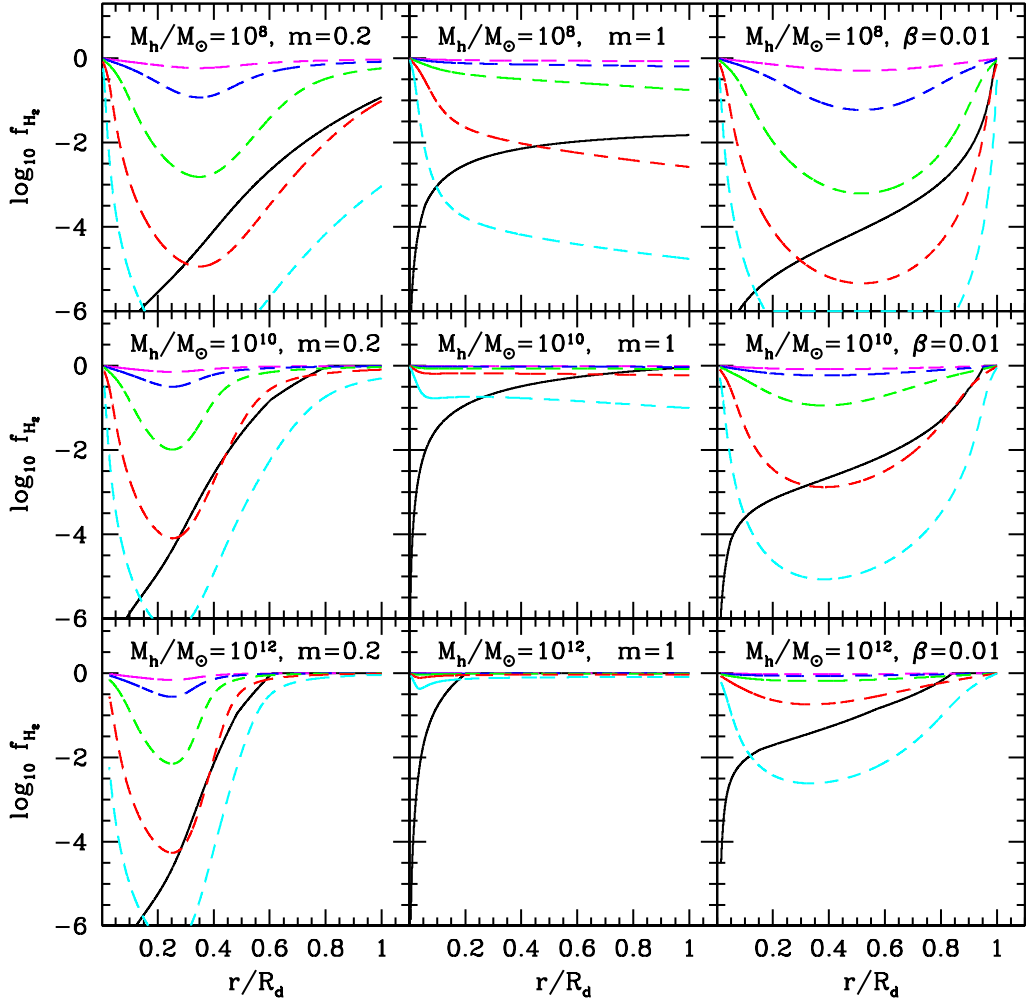}
\caption{\label{fig:fH2} 
The molecular fraction, $\fgmc$, as a function of galactocentric radius for our three fiducial disk models $z=6$.  Solid, black curves in the left, center, and right panels show our results from Eq. \ref{eq:fH2} for LSW models with $m=0.2$ and 1 and a nonlinear infall model with $\beta=0.01$, respectively.  We assume host halo masses of $10^{8}$ (top), $10^{10}$ (middle), and $10^{12}\,\msun$ (bottom).  Dashed curves show $\fgmc$ given by the model of \citet{Krumholz09b} in Eq. \ref{eq:fH2_krum} with $c=5$.  Each dashed line from top to bottom is calculated for $Z'=1$, 0.3, 0.1, 0.03, and 0.01.  
}
\end{center}
\end{figure*}

To calculate ${\rm SFR_{\rm ff}}/t_{\rm ff}$, we borrow from the analysis of \citet{KM05}, who derive  
\begin{equation}\label{eq:SFRff_tff}
\frac{{\rm SFR_{\rm ff}}}{t_{\rm ff}} \approx \frac{\epsilon_{\rm core}\,\alpha_{\rm vir}^{-0.68}\,\mach^{-0.32}\,M_{\rm H_{2}, 6}^{-1/4}\,{\Sigma}_{\rm cl, 85}^{3/4}}{53\,{\rm Myr}},
\end{equation}
where $M_{\rm H_{2}, 6}=\mgmc/10^6\,\msun$, and \citet{Krumholz09b} adopt $\epsilon_{\rm core}=0.3$ and a fixed virial ratio of $\alpha_{\rm vir}=2$.  The 1D thermal Mach number of the turbulence in the GMC is given by $\mach=\sigcl/c_{\rm s, th}$, where the cloud velocity dispersion is given by
\begin{equation}\label{eq:sigma_cl}
\sigcl=\left(\frac{\alpha_{\rm vir}\,G\,\mcl}{5\,\rcl}\right)^{1/2} \approx 3.75\,\kms\,\alpha_{\rm vir}^{1/2}\,M_{\rm cl, 6}^{1/4}\,\Sigma_{\rm cl, 85}^{1/4},
\end{equation}
and the thermal sound speed is $c_{\rm s, th}=\sqrt{\kb\,T/\mH}$.  Substituting for $\sigcl$ and $T$ from equation \ref{eq:temp}, the thermal Mach number is
\begin{equation}\label{eq:mach}
\mach \approx 11\,\alpha_{\rm vir}^{1/2}\,M_{\rm cl, 6}^{1/4}\,\Sigma_{\rm cl, 85}^{1/4}\,\dot{\Sigma}_{\rm \star, 1}^{-1/8}.
\end{equation}
These Mach numbers are, thus, typically supersonic.  Substituting into equation \ref{eq:SFRff_tff} with $\epsilon_{\rm core}=0.3$ and $\alpha_{\rm vir}=2$, we have
\begin{equation}\label{eq:SFRff_tff_2}
\frac{{\rm SFR_{\rm ff}}}{t_{\rm ff}} \approx \frac{\fgmc^{-1/4}\,M_{\rm cl, 6}^{-0.33}\,{\Sigma}_{\rm cl, 85}^{0.67}\,\dot{\Sigma}_{\rm \star, 1}^{0.04}}{680\,{\rm Myr}}.
\end{equation}

Substituting this expression into equation \ref{eq:SFlaw} and solving for $\fgmc$ gives
\begin{equation}\label{eq:fH2_2}
\fgmc \approx \left[\left(\frac{0.68\,{\rm Gyr}}{\Sgas/\dSstar}\right)\,M_{\rm cl, 6}^{0.33}\,\Sigma_{\rm cl, 85}^{-0.67}\,\dot{\Sigma}_{\rm \star, 1}^{-0.04}\right]^{1.33}.
\end{equation}
Equation \ref{eq:fH2_2} was derived with no concern for the physicality of $\fgmc$; if the star formation rate surface density at a given radius is higher than can be accounted for by the total gas present and the efficiency specified by equation \ref{eq:SFRff_tff}, then equation \ref{eq:fH2_2} will simply return a molecular fraction greater than unity.  The physical implication of this scenario is that the clouds need to be more efficient than predicted by equation \ref{eq:SFRff_tff_2} to produce a star formation rate capable of supporting $Q=1$ in the disk.  However, except at the edges of the galactic disks, our method typically produces results for $\fgmc$ in a physical range.  Nonetheless, to account for the possibility, we simply set $\fgmc$ to be the minimum of equation \ref{eq:fH2_2} and unity.  This additional condition effectively raises the efficiency of star formation from molecular gas such that the disk is self-regulated by star formation.  The result is the expression previously given in equation \ref{eq:fH2}.  

We plot the resulting molecular fraction in Figure \ref{fig:fH2} for our fiducial galaxies as a function of radius and compare it to the values obtained using the radiative and chemical equilibrium method of \citet{Krumholz09a}.  In their prescription, $\fgmc$ is an analytic function only of the surface density of the cold gas complex, $\Sigma_{\rm comp}=c\,\Sgas$, where $c$ is a clumping factor, and of metallicity:
\begin{equation}\label{eq:fH2_krum}
\fgmc \approx 1-\left[1+\left(\frac{3}{4}\,\frac{s}{1+\delta}\right)^{-5}\right]^{-1/5},
\end{equation}
where $s={\ln}(1+0.6\,\chi)/(0.04\,\Sigma_{\rm comp, 1}\,Z')$, $\chi=0.77\,(1+3.1\,{Z'}^{0.365})$, $\delta=0.0712\,(0.1\,s^{-1}+0.675)^{-2.8}$, and $\Sigma_{\rm comp, 1}=\Sigma_{\rm comp}/(1\,\msun\,{\rm pc^{-2}})$.\footnote{We have presented the slightly updated approximation given in \citet{Krumholz09b}, which is more accurate for small molecular fractions.  However, we opt not to use the further refinements of \citet{MK10}; though applicable to lower metallicities, this further method introduces additional dependences on the dust cross-section, the external radiation field, and the clump density---none of which can be usefully constrained in these galaxies.}  Since the clumping factor is sensitive to spatial resolution, it is difficult to determine observationally.  However, we follow \citet{Krumholz09b} in adopting $c=5$ for our comparison.  According to equation \ref{eq:fH2_krum}, the gas becomes fully molecular when $\Sgas \gtrsim 10/(c\,Z')\,{\rm \msun\,pc^{-2}}$.  By comparing with the top panel of Figure \ref{fig:sigma_sfr}, we can see that this should always be the case for galaxies hosted by $10^{10}\,\msun$ halos for each of our three fiducial disk models if $c\,Z'\gtrsim1$.  Figure \ref{fig:fH2} shows the full radial dependence of $\fgmc$ calculated using this method for the densities given by our galaxy models and a range of metallicities between $Z'=0.01$ and 1.  However, \citet{Krumholz09b} note that, while their general method is applicable to gas of arbitrary metallicity, the particular approximation of equation \ref{eq:fH2_krum} is not valid for $Z'<0.05$.  Thus, the extrapolations to lower metallicities should be evaluated accordingly.  

\begin{figure*}
\begin{center}
\includegraphics[width=6.0in,trim=0 332 0 0,clip]{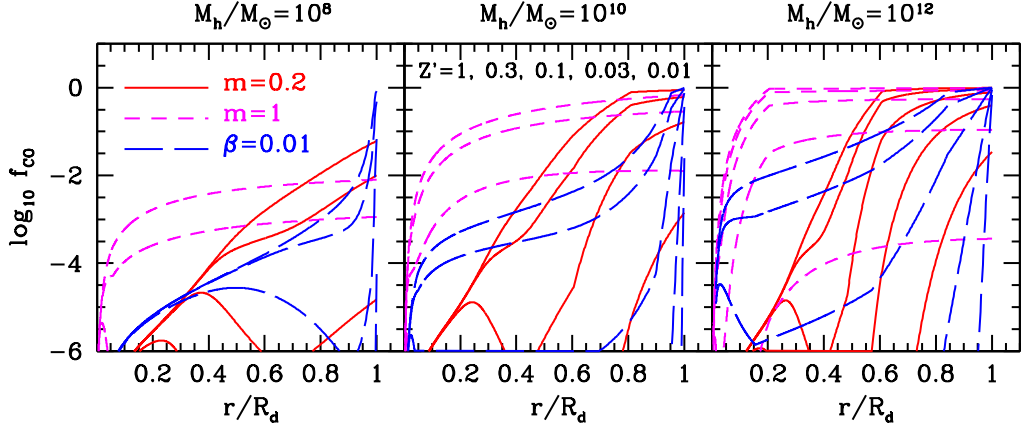}
\caption{\label{fig:fco} 
The fraction of all gas, $\fco$, in the central, CO-containing region of a cloud as a function of radius for our fiducial disk models at $z=6$ assuming host halos masses of $10^{8}$ (left), $10^{10}$ (center), and $10^{12}\,{\msun}$ (right).  As in Fig. \ref{fig:sigma_sfr}, solid (red), short-dashed (magenta), and long-dashed (blue) curves correspond to LSW models with $m=0.2$ and 1 and a nonlinear infall model with $\beta=0.01$, respectively.  For each disk model, lines from top to bottom were calculated for metallicities of $Z'=1$, $0.3$, $0.1$, $0.03$, and 0.01.  In cases where fewer than four lines of a given type are shown, only the highest metallicities appear on the plot.
}
\end{center}
\end{figure*}

Given its ease of use and transparent physical origins, equation \ref{eq:fH2_krum} is quickly becoming a common means of calculating the molecular fraction in semi-analytic and numerical models of galaxy formation \citep[e.g.][]{Narayanan12, Vallini12, Kuhlen12}.  However, our model does not provide a simple prediction for the gas-phase metallicity of $z\gtrsim6$ galaxies.  Observational determinations are difficult because of the complexity of and degeneracies within stellar population synthesis models in addition to unresolved nebular emission lines \citep[e.g.,][]{Dunlop12, deBarros12}.  Numerical simulations predict metallicities in the range of $Z' \approx 0.1-0.5$, but these may be over-predictions especially in the faintest galaxies \citep{Salvaterra11, Finlator11}.  On the other hand, while our calculation of $\fgmc$ does not depend directly on metallicity, our fitting to the UV luminosity function in \S\ref{sec:disk} \citep[see][]{Munoz12} may have encoded metallicity information.  Fixing $c=5$, we obtain rough agreement between our results and those using the \citet{Krumholz09b} method for $\fgmc$ by naively extrapolating down to $Z'\sim0.03$, somewhat more metal-poor than the predictions of \citet{Salvaterra11} and \citet{Finlator11}.  However, we can obtain the same agreement at $Z'\sim0.1$ if molecular clouds in high-redshift galaxies are somewhat less clumpy with $c\sim2$.  We leave a more thorough exploration of the implications of these results for the local GMC chemistry and structure to future work.  However, in the next section, the generally low value of $Z'$ will crucially limit the amount of carbon in the form of CO.

In our model here, we have considered only star formation in molecular gas.  However, star formation may proceed in atomic gas if the metallicity is low enough that thermal equilibrium is reached within a few cloud free-fall times but chemical equilibrium is not \citep{Krumholz12}.  While this scenario typically requires metallicities below a few percent of solar, i.e., at the very low end of what is predicted here for $z\gtrsim6$ galaxies, such circumstances may be more easily achieved for the decreased free-fall times of the higher density clouds at high-redshift.  If this effect is significant in the galaxies we consider, our method will somewhat overestimate the molecular fraction and, ultimately, the resulting CO luminosity.

%-------------------------------------------------------------------------------------------------------------
%             The Dark Fraction
%-------------------------------------------------------------------------------------------------------------
\subsection{The Dark Fraction}\label{sec:H2:fco}

In this section, we consider how much of the carbon in a cloud is actually in the form of CO.  Because H$_{2}$ self-shields more effectively than CO, carbon must rely more heavily on dust to maintain its molecular state.  This introduces a metallicity dependence into $\fco$, the fraction of the total cloud gas mass containing CO.  The fraction of molecular gas mass, thus, not probed by CO observations---often called the dark fraction---is $f_{\rm DF}=1-\fco/\fgmc$.  Because the mass enclosed within a given cloud radius is proportional to $r_{\rm cl}^2$, the radius within which carbon is primary in the form of CO is $\rco=\fco^{1/2}\,\rcl$.

In their PDR model, \citet{Wolfire10} assume that turbulently-generated inhomogeneities within clouds can be described as uniform clumps of density $n_{\rm c}$ embedded within more diffuse smooth density gas.  The clumps themselves are optically thin to external radiation but provide all of the opacity shielding the gas deeper within the cloud.  The resulting fraction of the gas containing CO is given by
\begin{equation}\label{eq:fco}
{\ln}\left(\frac{\fco}{\fgmc}\right)=\frac{-4.0}{A_{V}}\,\left[0.53-0.045\,{\ln}\left(\frac{G'_0}{n_{\rm c}\,{\rm cm^3}}\right)-0.097\,{\ln}(Z')\right],
\end{equation}
where $G'_0$ is the external radiation field in the far UV bathing the GMC in units of the local Galactic interstellar field found by \citet{Draine78}.  

As discussed in \citet{Wolfire10} and as is evident from equation \ref{eq:fco}, the dark fraction is relatively insensitive to changes in the ambient radiation field or the clump density and depends on metallicity primarily through its effect on $A_{V}$.  Therefore, we require only rough estimates of $G'_0$ and $n_{\rm c}$ to evaluate equation \ref{eq:fco}.  We approximate the external radiation field as $G'_0\approx 0.5\,\epsilon\,\Sigma_{\rm \star}\,c^2/(1.6\times10^{-3}\,{\rm erg/s/cm^2})$.\footnote{Interestingly, this reproduces empirical Kennicutt relation between star formation rate and FIR luminosity.}  We further assume $n_{\rm c}=\nmed$, where 
\begin{equation}\label{eq:nmed}
\nmed \approx \nave\,\sqrt{1+3\,\mach^2/4}
\end{equation}
is the mass-weighted median density in turbulent clouds \citep[see discussion and citations in \S\ref{sec:excitation} and][]{Wolfire10} and where $\mach$ is the 1D thermal Mach number given by equation \ref{eq:mach}.  We note that the higher radiation field expected in high-redshift galaxies affects the bracketed portion of equation \ref{eq:fco} in the opposite direction as the higher densities and lower metallicities.  

\begin{figure*}
\begin{center}
\includegraphics[width=6.0in,trim=0 332 0 0,clip]{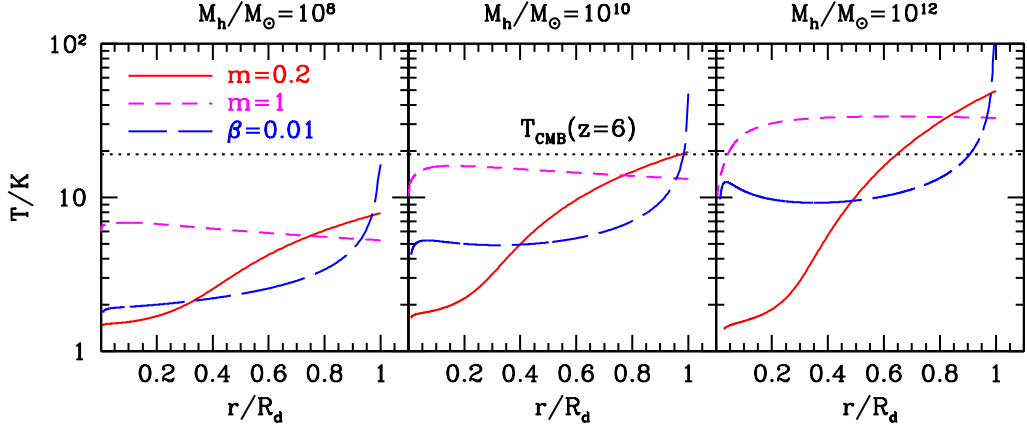}
\caption{\label{fig:temp} 
The local effective temperature of the disk, $T$, as a function of radius for our fiducial disk models at $z=6$ assuming host halos masses of $10^{8}$ (left), $10^{10}$ (center), and $10^{12}\,{\msun}$ (right).  As in Fig. \ref{fig:sigma_sfr}, solid (red), short-dashed (magenta), and long-dashed (blue) curves correspond to LSW models with $m=0.2$ and 1 and a nonlinear infall model with $\beta=0.01$, respectively.  For comparison, the horizontal, dashed line denotes the CMB temperature at $z=6$.  For calculating CO level populations and fluxes, we set the gas temperature to be $\tgas^4=T^4+\tcmb^4$.
}
\end{center}
\end{figure*}

In Figure \ref{fig:fco}, we plot $\fco$ as a function of radius for our three fiducial galaxy models and show the dependence on cloud metallicity.  Metallicity clearly has an exponential effect on the fraction of gas that contains CO.  However, the details depend on the ability of the hydrogen gas itself to shield the CO, i.e., for $A_{V}$ to be high despite low metallicity.  The recent work by \citet{Narayanan12} emphasizes the significant influence of a lowered metallicity on CO emission through this mechanism, but the treatment in \citet{Obreschkow09b} includes only a linear dependence on $Z'$.

%-------------------------------------------------------------------------------------------------------------
%             Gas Temperature 
%-------------------------------------------------------------------------------------------------------------
\section{Gas Temperature}\label{sec:temp}

The disk model in \S\ref{sec:disk} gives the effective temperature of the star forming disk in equation \ref{eq:temp}.  However, we also want to consider the effect on the gas of the incident CMB, whose temperature at high redshift is much higher than at $z=0$.  We assume that some combination of dust and line absorption allows the CMB to impart a significant fraction of its energy to the gas, energy that then contributes to the kinetic motions of all particles in the gas.  If the entire spectral energy of the CMB is absorbed in this way, the net effect is to raise the gas temperature to a value:
\begin{equation}\label{eq:tgas}
\tgas^4=T^4+\tcmb^4,
\end{equation}
where $\tcmb=19.11\,{\rm K}\,(1+z)/7$ is the temperature of the CMB and $T$ is given by equation \ref{eq:temp}.  In principle, this additional energy also contributes to the thermal pressure of the gas in equation \ref{eq:pgas}.  However, since we expect more CO emission from the larger area in the outer portions of the disk where pressure from stars dominates, we ignore this extra contribution to the thermal pressure for simplicity.  While equation \ref{eq:tgas} includes the effect of the CMB, we have neglected the contribution from cosmic rays.  This choice is also primarily for simplicity given the uncertainty in extrapolating their significance to $z\gtrsim6$ from low redshift determinations. In Figure \ref{fig:temp}, we plot the effective temperature of the disk as a function of fractional disk radius for galaxies in our model at $z=6$ and compare with the CMB temperature at the same redshift; the combination of these two temperatures determine the gas temperature through equation \ref{eq:tgas}.  Heating from the CMB will be particularly important for galaxies in low-mass halos but will not significantly affect the most UV-bright systems.

In contrast to our method in equation \ref{eq:tgas}, \citet{Krumholz11} provide a physical determination of $\tgas$ by considering gas in thermal equilibrium heated by cosmic rays and the photoelectric effect of dust grains and cooled by atomic and molecular line emission.  This is the method incorporated into the simulations of \citet{Narayanan12}.  However, their treatment does not account for heating from the CMB---which can be neglected relatively safely at $z\sim2$ (their definition of ``high-redshift")---and requires knowledge of several critical free-parameters, such as the dust-to-gas coupling coefficient and the cosmic ray energy density, which are not well-determined at higher redshift.  However, in both our analysis and that of \citet{Krumholz11} and \citet{Narayanan12}, the temperature is assumed not to vary within a cloud, primarily for simplicity.  In reality, the innermost regions of clouds will be somewhat cooler leading to a slight overestimate of our resulting CO emission.

%-------------------------------------------------------------------------------------------------------------
%             CO Excitation 
%-------------------------------------------------------------------------------------------------------------
\section{CO Excitation}\label{sec:excitation}

To calculate the optical depth and luminosity of CO transitions, we must consider the state of their level populations.  It is not valid to equate $\tgas$ with an excitation temperature if the level populations are not in thermal equilibrium.  In the rest of this work, we consider rotational transitions $J \rightarrow J-1$ having rest-frame frequency $\nuco\,J$, where $J$ is the rotational quantum number of the upper state and $\nuco=115.3\,{\rm GHz}$ is the frequency of the CO(1--0) transition.  In particular, we are interested in the $J=1$ transition---the most likely to be thermalized---and the $J=6$ transition, since its observed frequency at $z=6$ falls into the lowest energy ALMA band.

Because we have information about the structure of the clouds within our model galaxies, we can consider the level populations and optical depth of CO in each cloud directly.  We make the simplifying assumption that the cloud is uniform within a radius $\rco$ with a density, $\nco \equiv \bar{n}\left(< \rco\right)$, given by
\begin{equation}\label{eq:n_cl}
\nco = 3\,\ncl/(4\,\rco)\approx 40\,{\rm cm^{-3}}\,\Sigma_{\rm cl, 85}^{3/2}\,M_{\rm cl, 6}^{1/2}\,\fco^{1/2}.
\end{equation}
The optical depth of the $J \rightarrow J-1$ transition at the line center is 
\begin{equation}
\tau_{J}=\frac{g_J}{g_{J-1}}\,\frac{A_{J{\rm ,} J-1}\,\lambda^3_{J}}{4\,(2\,\pi)^{3/2}\,\sigcl}\,n_{J-1}\,\rco\,\left(1-\frac{n_{J}}{n_{J-1}}\,\frac{g_{J-1}}{g_J}\right),
\end{equation}
where $g_J=2\,J+1$ is the degeneracy of the $J$th state, $A_{J{\rm ,} J-1}$ is the Einstein--A emission coefficient for the transition, $\lambda_J=c/J\,\nuco$, and $n_J$ is the density of CO molecules in the $J$th state.  We take values of $A_{J{\rm ,} J-1}$ from the online data files of \citet{Moldata05} updated with recent results from \citet{Yang10}.\footnote{\label{fn:moldata}http://home.strw.leidenuniv.nl/~moldata/CO.html}  The density in the $J-1$ state is given by $n_{J-1}=[3\,\nco/4\,\rco]\,[n(\rm {CO})/n_{\rm H}]\,\pfunc_{{\rm CO,} J-1}$, where $n(\rm {CO})/n_{\rm H}\approx1.5\times10^{-4}\,Z'$ is the CO abundance of the gas and $\pfunc_{{\rm CO,} J-1}$ is the CO partition function in the $J-1$ state.  In thermal equilibrium, $n_{J}\,g_{J-1}/n_{J-1}\,g_{J}=\exp(-\hp\,J\,\nuco/\kb\,\tgas)$ and $\pfunc_{{\rm CO,} J-1}\approx \exp[-\hp\,J\,(J-1)\,\nuco/\kb\,\tgas]/\sqrt{1+(\kb\,\tgas/\hp\,\nuco)^2}$.  Because $\nco\,\rco$ is independent of $\fco$, the optical depth depends on metallicity only as $\tau_J \propto Z'$ in thermal equilibrium.  

Given an optical depth $\tau_J$ for the $J \rightarrow J-1$ transition, only a fraction $\beta_{J}$ of the line photons produced escape the cloud on average, where \citep{Draine11}
\begin{equation}\label{eq:beta}
\beta_{J}\approx \frac{1}{1+0.5\,\tau_J}.
\end{equation}
This trapping of photons decreases the critical density necessary for thermal excitation of a transition because radiative excitation will supplement collisional.  This effective critical density is given by
\begin{equation}\label{eq:ncrit}
\ncrit \approx \frac{\beta_J\,A_{J{\rm ,} J-1}}{k_{J{\rm ,} J-1}},
\end{equation}
where $k_{J{\rm ,} J-1}$ is the collisional de-excitation rate coefficient for the transition by H$_2$ taken from \citet[][see footnote \ref{fn:moldata}]{Moldata05}.  Emission from the CMB will have a similar effect on the level populations, particularly in the CO(1--0) transition and where the line is sub-thermally populated and optically thin (see Appendix \ref{sec:app:KT:cmb}).

In general, a given transition will be thermally populated if the gas density exceeds $\ncrit$ for that transition.  For a uniform cloud, all gas in the cloud fares the same in this respect.  However, fluctuations in the density within a cloud will change the average thermalization of the gas.  Because the dependence of the level populations on density below $\ncrit$ is non-linear, even Gaussian density fluctuations about the mean will have a non-zero effect on the cloud average.  In our picture of molecular clouds, the central portions of a cloud are shielded from dissociating radiation by turbulently-generated clumps whose densities can be much higher than the average.  A combination of observations and numerical tests show that supersonic, isothermal turbulence in local molecular clouds produces a log-normal density distribution with median density given by equation \ref{eq:nmed} \citep[see][and references therein]{KT07, Wolfire10}.  In very sub-thermal, highly supersonic regions, the gas in clumps can have a large effect on the expected emission.  However, if the amount of turbulence is not the dominant internal support for the cloud or the average density is high enough that the line of interest is already thermalized, then the influence of clumps on the expected emission will be small.

In practice, $\nco$, $\tau_{J}$, and $\mach$ span a wide range of values depending on metallicity, halo mass, and galactocentric radius.  Thus, certain lines will be thermal in some regimes but not in others.  Moreover, the determinations of level populations and optical depth are coupled due to photon trapping in thick regions.  Rather than commit to particular assumptions for optical depth and thermalization, we adopt the flexible approach provided by \citet{KT07}, which solves the coupled equations of detailed balance assuming the CO is in chemical (even if not necessarily {\emph{thermal}}) equilibrium.  We use a modified version of their publicly available code to calculate the populations, optical depths, and emissivities of each line as a function of cloud properties (see Appendix \ref{sec:app:KT}).\footnote{http://www.ucolick.org/~krumholz/downloads.html}  The code accounts for the log-normal distribution of gas densities within a cloud generated by turbulence by computing the level populations as a function of density.  Of course, this code has its own built-in assumptions and simplifications, such as assuming a constant density throughout the cloud when calculating escape probabilities, but it currently remains the best available approach for addressing this problem.

\begin{figure*}
\begin{center}
\includegraphics[width=6.0in,trim=0 166 0 0,clip]{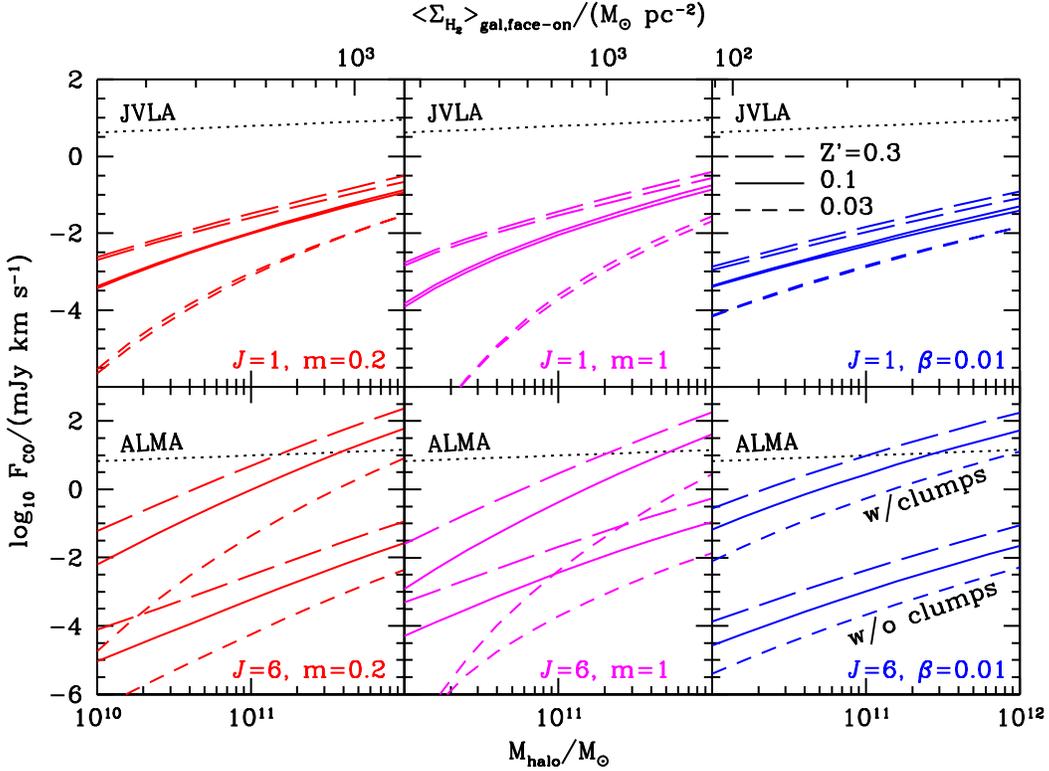}
\caption{\label{fig:FVco} 
The CO(1--0) (top panels) and CO(6--5) (bottom panels) flux as a function of halo mass for our fiducial disk models at $z=6$.  The top axis shows the average molecular surface density for a face-on galaxy corresponding to each halo mass.  Long-dashed, solid, and short-dashed curves assume $Z'=0.3$, 0.1, and 0.03, respectively.  Upper lines of a given type include turbulent clumps in the calculation of level populations as described in \S\ref{sec:excitation}, while we have ignored these fluctuations in the lower lines.  The noise levels in ten hour observations with JVLA and ALMA are denoted by the dotted lines.  As in Fig. \ref{fig:fH2}, the left, center, and right panels show results for LSW models with $m=0.2$ and 1 and a nonlinear infall model with $\beta=0.01$, respectively.
}
\end{center}
\end{figure*}

%-------------------------------------------------------------------------------------------------------------
%             CO Luminosity
%-------------------------------------------------------------------------------------------------------------
\section{CO Luminosity}\label{sec:Lco}

%-------------------------------------------------------------------------------------------------------------
\subsection{Emission from Molecular Clouds}\label{sec:Lco:clouds}

The \citet{KT07} code gives the frequency-integrated luminosity density, $L'_J$, for each line.  The {\emph{velocity}}-integrated luminosity of the cloud is given by $(4\,\pi/3)\,\rco^3\,c\,L'_J/(J\,\nuco)$.  While clouds at the outskirts of the disk have lower emissivities than those near the center of the disk, they also have much larger regions that contain CO.  By combining this cloud luminosity with the number of clouds as a function of galactocentric radius (Eq. \ref{eq:N}), our model predicts a radial profile of CO flux in which the majority of emission is not concentrated at the center but rather produced in the outer portions of the disk.   

We can obtain the total luminosity escaping GMCs from each radius of the galaxy by simply counting the number of clouds at each disk radius:
\begin{equation}\label{eq:Lco}
\frac{dL_{J}}{d{\ln}r}=\frac{4\,\pi\,\rco^3\,c\,L'_J}{3\,J\,\nuco}\,\frac{dN}{d{\ln}r}.
\end{equation}
Integrating equation \ref{eq:Lco} over radius will give the total luminosity, $L_{J}$, for the $J$th transition emitted by clouds in the galaxy.  

If the transitions of interest were in thermal equilibrium and optically thick, their line luminosities would be straightforward to calculate.  In this case, the velocity-integrated luminosity emitted from the surface at $\rco$ by each cloud is $8\,\pi^2\,\rco^2\,\sigcl\,B(J\,\nuco, \tgas)$, where we have assumed that $B(J\,\nuco, \tgas)$, the Planck function evaluated at frequency $J\,\nuco$ and with brightness temperature $\tgas$, is constant over a top-hat line profile with a turbulent width given by $2\,\sigcl$.  In the Rayleigh--Jeans limit, $B(J\,\nuco, \tgas) \propto J^2\,\tgas$, and, if $\tgas$ is dominated by $\tcmb$, then equation \ref{eq:Lco} depends on disk, GMC, and line properties as $dL_J/d{\ln}r \propto J^2\,\sigma^{-1}\,\fco^{5/4}\,\Sgas^{7/4}\,\Scl^{-3/4}\,r^3$.  For $\Sgas>85\,{\rm \msun/pc^2}$, this reduces to $dL_J/d{\ln}r \propto J^2\,\sigma^{-1}\,\fco^{5/4}\,\Sgas\,r^3$.  The remaining density dependence in $\fco$ means that $L_{J}$ is not proportional to $\Sgas$ as is often assumed under these assumptions.  The dependence of $L_{J}$ on $\fco$, moreover, makes it a strong function of metallicity.

We use relations found in Appendix A of \citet{Obreschkow09b} to convert the CO luminosities we calculate into quantities convenient for observers.  The observed, velocity-integrated flux from each galaxy is 
\begin{equation}\label{eq:Fco}
\Fco=\frac{(1+z)\,L_{J}}{4\,\pi\,D_{\rm L}^2},
\end{equation}
where $D_{\rm L}$ is the luminosity distance of the galaxy at redshift $z$.  If the observing beam size is larger than the galaxy on the sky, then the flux in equation \ref{eq:Fco} can be expressed as a temperature as
\begin{equation}\label{eq:Tco}
T_{\rm CO}=\frac{c^2\,(1+z)^3}{2\,\kb\,J^2\,\nuco^2\,\Omega_{\rm gal}}\,\Fco,
\end{equation}
where $\Omega_{\rm gal}$ is the solid angle of the galaxy on the sky, and $\Omega_{\rm gal}=\pi\,\rdisk^2/D_{\rm A}^2$ for a face-on galaxy, where $D_{\rm A}$ is the angular diameter distance to the galaxy at redshift $z$.

\begin{figure}
\begin{center}
\includegraphics[width=\columnwidth,trim=0 0 230 0,clip]{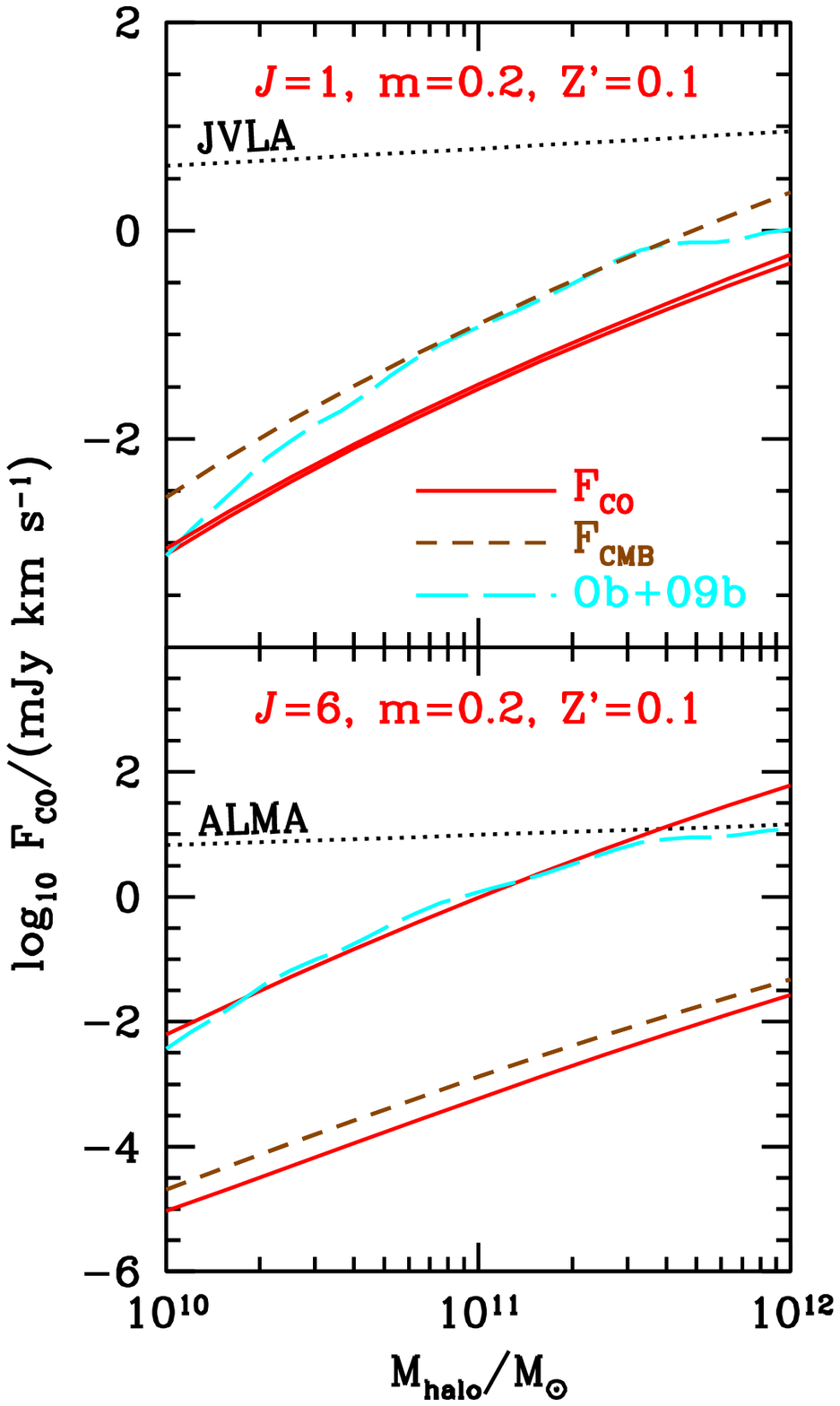}
\caption{\label{fig:Fco_comp} 
The CO(1--0) (top panel) and CO(6--5) (bottom panel) flux as a function of host halo mass at $z=6$.  Solid lines show the integrated flux produced by clouds in the galaxy, while short-dashed lines denote the required CMB flux to be subtracted (Eq. \ref{eq:Fcmb}) assuming $Z'=0.1$ and an LSW disk model with $m=0.2$.  As in Fig. \ref{fig:FVco}, results are plotted both including (upper solid lines) and excluding (lower solid lines) the effect of turbulent clumps on the level populations.  The mean CMB-subtracted CO emission from \citet{Obreschkow09b} is plotted in the long-dashed curves (see discussion in \S\ref{sec:comp}).  Dotted lines denote the noise levels in ten hour observations with JVLA and ALMA.
}
\end{center}
\end{figure}

%-------------------------------------------------------------------------------------------------------------
\subsection{CMB Subtraction}\label{sec:Lco:cmb}

\citet{Obreschkow09b} and \citet{daCunha13b} emphasized that the CMB at high redshift should be considered, not just as a heating term, but as an observational background.  To estimate the subtraction of this background, first consider the contributions to the total flux observed from a galaxy with solid angle $\Omega_{\rm gal}$ on the sky.  At the frequency of the CO line of interest, emission is observed from GMCs within the galaxy and from the CMB, which we assume passes freely between clouds but may be absorbed if transmitted through clouds.  Let $\Omega_{{\rm cl,} i}$, $I_{{\rm cl,} i}$, and $x_{i}$ denote the solid angle, intensity, and fractional CMB transmission of the $i$th cloud numbered 1 to $N$.  Assuming non-overlapping clouds, the total flux observed from the galaxy is
$$\sum\limits_{i=1}^N \left(I_{{\rm cl,} i}+x_{i}\,I_{\rm CMB}\right)\,\Omega_{{\rm cl,} i}+I_{\rm CMB}\,\left(\Omega_{\rm gal}-\sum\limits_{i=1}^N \Omega_{{\rm cl,} i}\right).$$  On the other hand, the CMB background from an equivalently sized area on the sky is
$$I_{\rm CMB}\,\Omega_{\rm gal}.$$
Assuming a fraction $x_{i}={\rm e}^{-2\,\tau_{J{\rm ,} i}}$ of CMB photons are transmitted through cloud $i$ with optical depth $\tau_{J{\rm ,} i}$ for the $J$th transition of CO, then the background subtracted flux is
\begin{eqnarray}
F_{\rm CO-CMB}\!&\!=\!&\!\sum\limits_{i=1}^N \left[I_{{\rm cl,} i}-\left(1-{\rm e}^{-2\,\tau_{J{\rm ,} i}}\right)\,I_{\rm CMB}\right]\,\Omega_{{\rm cl,} i}\nonumber\\
\!&\!=\!&\!\Fco-\Fcmb,
\end{eqnarray}
where
\begin{equation}\label{eq:Fcmb}
\Fcmb=8\,\pi^2\,B(J\,\nuco, \tcmb)\,\int\!\! \rco^2\,\sigcl\,\left[1-{\rm e}^{-2\,\tau_{J}(r)}\right]\,\frac{dN}{d{\ln}r}\,{d{\ln}r}.
\end{equation}
Thus, the amount of CMB subtraction depends not just on the CMB itself but on the size and transparency of the clouds.  A positive value of $F_{\rm CO-CMB}$ corresponds to CO seen in emission, while a negative value indicates absorption.

%-------------------------------------------------------------------------------------------------------------
\subsection{Predicted CO Fluxes}\label{sec:Lco:results}

Ultimately, we want to compare the resulting emission to the observational thresholds of ALMA and the JVLA.  Observing at $30\,{\rm GHz}$ in a $200\,\kms$ channel, the JVLA reaches a noise level of $42\,{\rm {\mu}Jy}$ after ten hours of observation.  If we assume that the line profile emitted by the galaxy has a width of $2\,\sigma$ dominated by the halo velocity dispersion, then the noise level on the velocity-integrated CO flux is $11.5\,{\rm mJy\,\kms}\,[J\,\sigma_{50}/(1+z)]^{1/2}$.  In comparison, the ALMA sensitivity per $200\,\kms$ channel at $90\,{\rm GHz}$ in a typical ten hour observation is $47\,{\rm {\mu}Jy}$, leading to a velocity-integrated noise level of $7.5\,{\rm mJy\,\kms}\,[J\,\sigma_{50}/(1+z)]^{1/2}$.

Our methodology determines $\Fco$ and $\Fcmb$ individually to sufficient accuracy for our purposes but often predicts that the difference between the two should be much smaller than each of their absolute values.  Consequently, $F_{\rm CO-CMB}$ is very sensitive to the simplifying assumptions and uncertain details of our calculation, even up to the sign of the result.  Therefore, in this section, we will show results only for $\Fco$ and $\Fcmb$ separately.  These ``raw" fluxes a clearer prediction of the model than is $F_{\rm CO-CMB}$, but we emphasize that real observations require CMB subtraction, and will present results using the CMB-subtracted signal in \S\ref{sec:results}.

Figure \ref{fig:FVco} shows the average $\Fco$ for $J=1$ and $J=6$ flux emitted by each of our model galaxies as a function of its host halo mass at $z=6$.  We plot results for $Z'=0.3$, 0.1, and 0.03 to cover the range of expected metallicities for these systems, though we remind the reader that our model is consistent with chemical equilibrium calculations at the lower end of this range (see \S\ref{sec:H2:fH2}).  Note that, in this work, we plot only the {\emph{average}} flux of galaxies hosted by the given halo mass and ignore fluctuations about that mean.  All of our different fiducial angular momentum transport models result in roughly similar CO luminosities, which indicates the robustness of our results.  We compare the effect of ignoring turbulent clumps within clouds in our calculation of the CO level populations.  We find that these inhomogeneities can have a big influence on emission in the $J=6$ line, whose critical density is typically much higher than the average cloud density.  On the other hand, since $\nco$ is rarely lower than the critical density for the $J=1$ line, an increase in the median density resulting from clumps does not significantly affect the thermalization.  

We robustly predict that the $J=1$ line in $z=6$ LBGs will be unobservable by JVLA after ten hours of observation.  However, the observability of the $J=6$ line with ALMA depends on several effects.  We find that the line may only be observable in the most massive halos and only on the conditions that $Z'\gtrsim0.1$ and that turbulent clumps impact the line thermalization to the extent that we have assumed.  For example, if turbulence is not the dominant source of support for high-redshift GMCs, then the clouds may be more homogenous than we predict and the average $J=6$ emission reduced.

The behavior with increasing redshift is complicated by a competition between several factors.  At fixed halo mass, galaxies become smaller and denser and their accretion rates from cold-flows increase.  However, these systems are also rarer, more distant, and potentially more metal-poor.  Moreover, the increasing temperature of the CMB provides additional heating, a greater influence on the level populations, and a higher observational background (\S\ref{sec:Lco:cmb}).  We leave a more detailed investigation of the subtleties of this redshift evolution to future work but note that the overall effect of increasing the redshift from $z=6$ to $z=7$ or 8 is minimal---no more than a factor of a few in either direction---and does not change the qualitative conclusion that the CO emission will be beyond the reach of modern facilities.

Figure \ref{fig:FVco} clearly shows the nonlinear dependence of $\Fco$ on $Z'$ as expected from the exponential relation in equation \ref{eq:fco}.  However, in the shocked infall model, most of the emission comes from the extreme outer portions of the disk where the cloud column densities are higher and the effect of a lowered metallicity is less extreme (see Fig. \ref{fig:fco}).  As a result, $\Fco$ is somewhat less sensitive to metallicity in this model.

In Figure \ref{fig:Fco_comp}, we compare $\Fco$ and $\Fcmb$ assuming $Z'=0.1$ and a fiducial LSW angular momentum transport model with $m=0.2$.  The figure shows that $\Fco$ is comparable to---or even a bit lower than---$\Fcmb$ for $J=1$ and, if we ignore turbulent clumps in our calculation of the level populations, for $J=6$.  Taken at face value, these calculations imply that arbitrarily sensitive measurements would observe the lines in these cases in absorption against the CMB background.  Accounting for the log-normal distribution of gas densities within the cloud as discussed in \S\ref{sec:excitation}, the $J=6$ line appears in emission with $\Fco>>\Fcmb$.

%-------------------------------------------------------------------------------------------------------------
%       X-Factor
%-------------------------------------------------------------------------------------------------------------
\section{The X-Factor}\label{sec:Xco}

\begin{figure*}
\begin{center}
\includegraphics[width=6.0in,trim=0 260 0 0,clip]{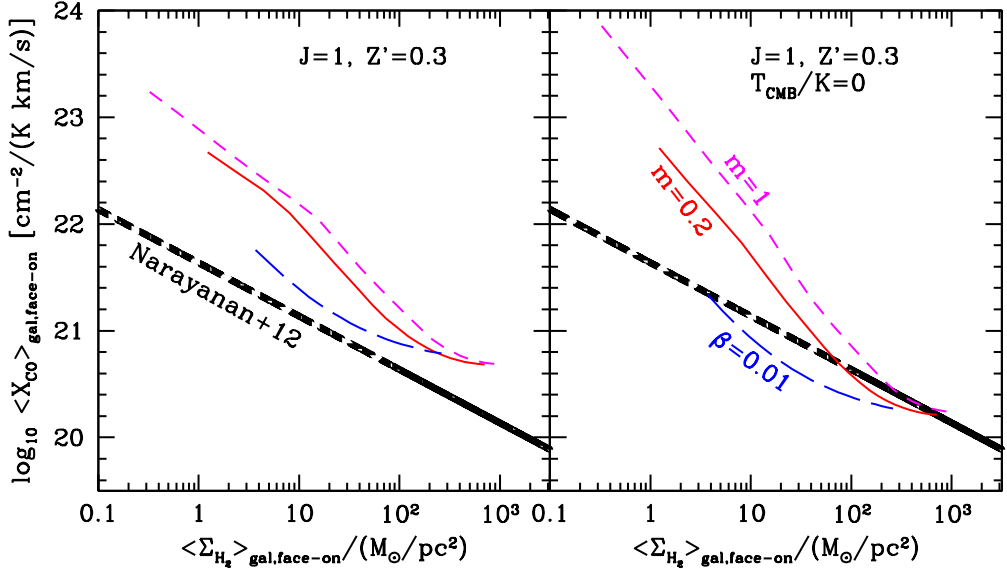}
\caption{\label{fig:Xco} 
The X-factor as a function of average molecular fraction for face-on galaxies in our fiducial disk models at $z=6$ with a metallicity of $Z'=0.3$.  In the left panel, we show results that include the CMB as a source of both heating and (de-)excitation.  Thin lines correspond to results using our methodology with solid (red), short-dashed (magenta), and long-dashed (blue) curves denoting LSW models with $m=0.2$ and 1 and a nonlinear infall model with $\beta=0.01$, respectively, as in Fig. \ref{fig:sigma_sfr}.  Thick curves were calculated using the \citet{Narayanan12} model (Eq. \ref{eq:xco}) but using our molecular surface densities; dashed portions indicate where we have extrapolated the fitting formula to surfaces densities below those actually simulated.  We also note that our definition of molecular density---calculated as $M_{\rm H_{2}, tot}/A$, where $M_{\rm H_{2}, tot}$ is the total molecular gas mass and $A$ is the face-on area of the galaxy---differs from the mass-weighted molecular densities of \citet{Narayanan12}.  In the left panel, we show our results including the CMB as a source in both heating and (de-)excitation (but not as an observational background), while in the right panel, we have repeated the calculation setting $\tcmb=0$ to compensate for the lack of a significant CMB in the \citet{Narayanan12} model.
}
\end{center}
\end{figure*}

The CO luminosity is often evaluated by its predicted relationship to gas density through the quantity $\xco$---the ratio of molecular gas surface density to the $J=1$ CO emission temperature---commonly referred to as the X-factor.  \citet{Narayanan12} used numerical simulations to predict a general expression for $\xco$ in terms of the average surface density of molecular gas and metallicity given by
\begin{equation}\label{eq:xco}
\xco \approx 1.3\times10^{21}\,{\rm \frac{cm^{-2}}{\K\,\kms}}\,{Z'}^{-1}\,\left(\frac{\left<\Sigma_{\rm H_2}\right>}{\rm \msun\,pc^{-2}}\right)^{-0.5},
\end{equation}
where $\left<\Sigma_{\rm H_2}\right>$ is the {\emph{mass-weighted}} surface density of molecular gas averaged over the galaxy.  In the left panel of Figure \ref{fig:Xco}, we compare this prediction to the intrinsic emission (neglecting the CMB background) from our model galaxies, taking $Z'=0.3$ as a fiducial case.  Note, however, that we define $\left<\Sigma_{\rm H_2}\right>=M_{\rm H_{2}, tot}/A$ for our results, where $M_{\rm H_{2}, tot}$ is the total molecular gas mass and $A$ is the face-on area of the galaxy.  Each point along a plotted line of our results corresponds to a different halo mass from $10^{8}\,\msun$ to $10^{12}\,\msun$ at $z=6$.  While the \citet{Narayanan12} model is only fit to simulated galaxies for $\left<\Sigma_{\rm H_2}\right>\gtrsim 100\,{\rm \msun/pc^2}$, we extrapolate their results via equation \ref{eq:xco} below the regime in which they tested it.  The agreement between our results and equation \ref{eq:xco} is quite good given our very different methodology.  

In the right panel of Figure \ref{fig:Xco}, we repeat the comparison with our model recalculated to exclude the CMB as a source of both heating and (de-)excitation since it is not included in the \citet{Narayanan12} model.  Correcting for this difference, the collective agreement of our results improves for the largest galaxies that we consider.  While there is still some disagreement in the smallest galaxies, we note both that equation \ref{eq:xco} is likely not valid at these low densities and that galaxies in halos as small as these may not be able to form anyway \citep[e.g.,][]{ML11}.  The demonstrated agreement for the largest galaxies---producers of the brightest CO emission, the most likely to exist, and the ones for which equation \ref{eq:xco} is less of an extrapolation---validates our approach.  In particular, this consistency suggests that our steady-state disk model is not an unreasonable choice for the high-redshift LBGs.

%-------------------------------------------------------------------------------------------------------------
%       Observational Strategies
%-------------------------------------------------------------------------------------------------------------
\section{Observational Strategies}\label{sec:results}

In \S\ref{sec:Lco} we showed that the CO emission produced by clouds in $z\gtrsim6$ LBGs will be very difficult to detect even for ALMA.  Moreover, the CMB-subtracted signal, $F_{\rm CO-CMB}$, is sensitive to the details of the calculation because we expect $\Fco$ and $\Fcmb$ to be close in value.  Nevertheless, in this section, we present CMB-subtracted results from two different strategies for observing CO lines at $z\gtrsim6$: blind searches and followup of UV-selected galaxies.  We calculate both the CO luminosity function and the expected CO flux as a function of UV magnitude.  This exercise demonstrates the functionality of our model, and the methods can be used in the future to predict observations for other chemical lines.

%-------------------------------------------------------------------------------------------------------------
\subsection{Surveys for Galaxies in CO}\label{sec:results:COLF}

\begin{figure}
\begin{center}
\includegraphics[width=\columnwidth]{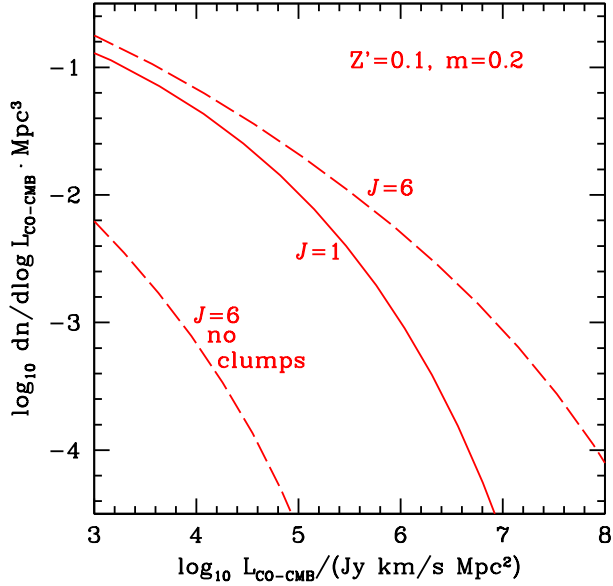}
\caption{\label{fig:COLF} 
The CO luminosity function at $z=6$.  Solid and dashed curves show results for the $J=1$ and $J=6$ transitions, respectively, assuming $Z'=0.1$ and a fiducial LSW disk with $m=0.2$.  For $J=6$, we compare results including (upper) and excluding (lower) the effect of turbulent clumps on the level populations.
}
\end{center}
\end{figure}

The ALMA field-of-view for observing CO transitions is approximately $64\,(6/J)\,[(1+z)/7]$ arcsec on a side, which can tile the $4.6\,{\rm arcmin^2}$ WFC3 IR field-of-view in about four pointings.  It will be interesting to compare the luminosity function of molecular line emission at high-redshift to those compiled from rest-frame UV data.  

After calculating the average CMB-subtracted CO luminosity, $L_{\rm CO-CMB}$, as a function of halo mass, we use the \citet{ST02} halo mass function to produce the CO luminosity function.  In principle, this calculation should include the scatter in luminosity for a given halo mass that results from variations in the galactic accretion rate from galaxy-to-galaxy.  These fluctuations serve to flatten the luminosity function just as it does in the UV \citep{ML11, Munoz12}.  However, in this work, we simplify the treatment and calculate the luminosity function by assuming a single value of the luminosity for each halo mass.  For further simplicity, we also ignore any cut-off of galaxy formation in low-mass halos.

We show results for the CO luminosity function at $z=6$ in Figure \ref{fig:COLF} assuming $Z'=0.1$ and a LSW angular momentum transport model with $m=0.2$, but, as shown in Figure \ref{fig:FVco}, different disk models would yield similar results.  We present the calculation for both the $J=1$ and $J=6$ transitions, considering for the latter the effect of both include and excluding turbulent clumps in the level population calculation.  As shown, inhomogeneities within clouds strongly influences the $J=6$ luminosity function since a moderate change in corresponding halo mass at each luminosity translates into a large change in abundance at $z=6$.  Additionally, as can be understood from Fig. \ref{fig:Fco_comp}, the clumps produces a signal seen in emission against the CMB background, while the $J=1$ signal and the $J=6$ signal without clumps are seen in absorption.  Finally, as demonstrated in Figure \ref{fig:FVco}, metallicity may also have a significant affect on the results; a small drop in metallicity may significantly lower the expected abundance at a given luminosity, which, as a result, can only be generated by a somewhat larger galaxy.

%-------------------------------------------------------------------------------------------------------------
\subsection{Followup of Existing LBG Candidates}\label{sec:results:CO-UV}

While telescopes like ALMA will likely not be as efficient as WFC3 in discovering new, large samples of galaxy candidates at $z\gtrsim6$ based on their CO line emission, they may be useful for the followup of UV-selected targets.  In this section, we calculate the expected CO flux as a function of $\muv$, the absolute AB magnitude in the rest-frame UV.  As with the luminosity function, the scatter in both CO luminosity and $\muv$ at fixed host halo mass enters into the calculation \citep{ML11, Munoz12, MF12}.  These fluctuations are correlated and lead to less scatter in CO flux at fixed $\muv$ than at fixed host halo mass.  However, given the difficulty in observing CO, we again perform a simplified computation in which every halo hosts a galaxy of the same average $\muv$ and $F_{\rm CO-CMB}$.

\begin{figure}
\begin{center}
\includegraphics[width=\columnwidth,trim=0 0 30 0,clip]{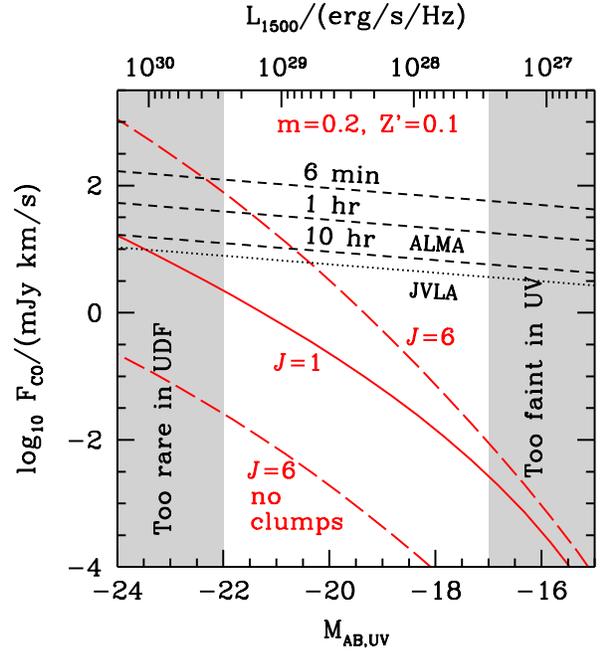}
\caption{\label{fig:Fco_vs_Muv} 
CO flux as a function of absolute UV magnitude at $z=6$.  As in Fig. \ref{fig:COLF}, solid and long-dashed curves show results for the $J=1$ and $J=6$ transitions, respectively, assuming $Z'=0.1$ and a fiducial LSW disk model with $m=0.2$.  Moreover, for $J=6$, we compare results including (upper) and excluding (lower) the effect of turbulent clumps on the level populations as in Fig. \ref{fig:COLF}.  The dotted line marks the noise level for JVLA observations of the $J=1$ transition integrated for 10 hrs, while the short-dashed lines indicate noise levels for ALMA observations of the $J=6$ transition integrated for 10 hrs, 1 hr, and 6 min.  Shaded regions indicate galaxies that would either be too faint in the UV for detection in UDF12 or too rare to be found in a UDF-sized area on the sky.  
}
\end{center}
\end{figure}

Figure \ref{fig:Fco_vs_Muv} shows the average CO flux expected from $z=6$ galaxies as a function of absolute magnitude in the rest-frame UV assuming $Z'=0.1$ and a LSW angular momentum transport model with $m=0.2$.  Again, as in Figure \ref{fig:FVco}, different disk models yield similar results, while varying the metallicity may have a stronger effect.  We have compared these results to the noise levels for measurements of the $J=1$ and $J=6$ transitions with JVLA and ALMA and to the approximate current detection limits in the UDF12 UV luminosity function \citep{Schenker13, McLure13}.  Clearly, galaxies are more easily found in the UV than in CO.  Only the most UV-bright galaxies may have detectable CO counterparts after reasonable integration times and only if both $Z'\gtrsim0.1$ and the effect of turbulent clumps is as significant as we have assumed.  Yet, these objects are also the rarest on the sky---no objects at $z=6$ brighter than $\muv\sim-22$ in the UV are found in the UDF \citep{Bouwens06}---providing an obstacle to the stacking of large samples.  Observation time would, therefore, be best spent integrating in a single pointing.

%-------------------------------------------------------------------------------------------------------------
%       Comparison with Previous Work
%-------------------------------------------------------------------------------------------------------------
\section{Comparison with Previous Work}\label{sec:comp}

In this section, we discuss the relationship between our work and recent efforts by \citet{Obreschkow09a} and \citet{Lagos12}.  

\citet{Obreschkow09a} modeled the CO luminosity by combining a semi-analytic model for galaxy formation \citep{Croton06, dLB07} with simple prescriptions for molecular emission extrapolated from local empirical relations.  Our formalism reproduces the normalization constant for their CO luminosity under a simple set of assumptions but predicts a much lower value under the physical conditions expected in the high-redshift ISM.

\citet{Obreschkow09b} give the CMB-subtracted luminosity as a function of molecular mass, metallicity, and excitation temperature as
\begin{equation}\label{eq:Ob}
L_J = M_{\rm H_{2}, tot} \, \frac{k\,Z\,\epsilon'}{B'} \, \left[1-{\exp}\left(-B'\,\tau_J\right)\right]\,l(J,\texc,z),
\end{equation}
where
\begin{equation}\label{eq:lOb}
l(J,\texc,z) \equiv \frac{J^4}{{\exp}\left(\frac{h_{\rm p}\,\nuco\,J}{\kb\,\texc}\right)-1}-\frac{J^4}{{\exp}\left(\frac{h_{\rm p}\,\nuco\,J}{\kb\,\tcmb}\right)-1},
\end{equation}
\begin{equation}\label{eq:tauOb}
\tau_{J}=7.2\,\tau_{\rm c}\,{\exp}\left(-\frac{h_{\rm p}\,\nuco\,J^2}{2\,\kb\,\texc}\right)\,{\rm sinh}\left(\frac{h_{\rm p}\,\nuco\,J}{2\,\kb\,\texc}\right),
\end{equation}
and $Z$ is the mass fraction of metals in the cold gas phase.  Additionally, $k=8\times10^{-8}\,{\rm W\,kg^{-1}}$ and $\tau_{\rm c}=2$ are determined using fits to the CO luminosity function with $Z\approx0.01$ (corresponding approximately to $Z'\approx1$) and spectral energy distributions of local galaxies.  $B'$ and $\epsilon'$ describe the degree of cloud overlap and the amount of smooth molecular gas not in clouds, respectively.  \citet{Obreschkow09b}, thus, assumes that the CO line emission is well-approximated by the emission from thermalized gas at a slightly under-estimated optical depth or temperature and expresses the excitation temperature of the CO in their model galaxies as
\begin{equation}\label{eq:Tob}
\texc^4=\tcmb^4+T_{0}^4+T_{\rm SB}^4+T_{\rm AGN}^4,
\end{equation}
where $T_{0}$, $T_{\rm SB}$, and $T_{\rm AGN}$ are the contributions from clump self-heating, star formation, and AGN, respectively.  The merging of relevant temperatures in equation \ref{eq:Tob} is similar to the way we combine the temperature of the disk with that of the CMB in equation \ref{eq:tgas}.  Since their semi-analytic models don't provide information about the radial distribution of a galaxy's properties, all quantities are averages over whole galaxies.  Assuming forms for the dependences of each term and calibrating to local systems, \citet{Obreschkow09b} set $T_{0}=17\,\K$ and 
\begin{equation}\label{eq:TSB}
T_{\rm SB}=60\,{\rm K}\,\left[\dSstar/(\dSstar+500\,{\rm \msun\,yr^{-1}\,kpc^{-2}})\right]^{1/4}.
\end{equation}
In the limit of $\dSstar \ll 500\,{\rm \msun\,yr^{-1}\,kpc^{-2}}$, equation \ref{eq:TSB} is nearly identical to equation \ref{eq:temp}.  Considering the hosts of $z\gtrsim6$ LBGs specifically, a $10^{10}\,\msun$ halo has an average star formation rate of about $0.31\,{\rm \msun/yr}$ and a disk radius of $0.36\,{\rm kpc}$ resulting in a face-on star formation surface density of $0.8\,{\rm \msun\,yr^{-1}\,kpc^{-2}}$ and $T_{\rm SB}\approx0.2$.  For $10^{11}$ and $10^{12}\,\msun$ halos, this rises to 1.5 and $2.2\,{\rm K}$, respectively.  Thus, $T_{\rm SB}$ makes only a small contribution to equation \ref{eq:Tob}.  Further, owing to the low black hole accretion rates of typical $z\gtrsim6$ galaxies \citep{MF12}, the contribution from AGN is negligible. 

Our methodology also differs from that of \citet{Obreschkow09b} in the calculation of molecular gas masses.  Our method, described in \S\ref{sec:H2:fH2}, is consistent with the \citet{Krumholz09a} model, which argues that density is the most physical quantity affecting the molecular fraction.  In contrast, \citet{Obreschkow09b}, along with several other studies \citep[e.g.,][]{Lagos12}, rely on the relationship between molecular fraction and mid-plane pressure empirically determined from local galaxies.  This relation gives a ratio of molecular to atomic gas of
\begin{equation}\label{eq:Rmol-P}
R_{\rm mol} = \left(\frac{P}{P_{\star}}\right)^{\alpha},
\end{equation}
where $R_{\rm mol}=\left(\fgmc^{-1}-1\right)^{-1}$, $P$ is the mid-plane pressure, $\alpha=0.8$, and $P_{\star}=2.35\times10^{-13}\,{\rm Pa}$ \citep{Leroy08}.  Molecular fractions were determined from CO(1-0) measurements with an assumed value of the X-factor.  \citet{Obreschkow09a} further assume exponential disk density profiles in their model galaxies and a relationship between this density and mid-plane pressure.  While this treatment may be sufficient for their calculations at $z=0$, understanding the underlying physical process responsible for the atomic to molecular transition is critical if the relations are extrapolated to environments very different from the ones in which they were calibrated.  Using the mid-plane pressure in our disk model (Eq. \ref{eq:pressure}), we find that the molecular fraction at $z=6$ calculated from equation \ref{eq:Rmol-P} is essentially unity at all radii.  That is, the high densities and thin disks of our galaxies result in gas that is completely molecular for all models we consider.  With molecular fractions this high, the star formation efficiencies we calculate in GMCs would imply star formation rates and UV luminosities somewhat higher than observed.  Given the relationship between molecular mass and CO luminosity, our lower molecular fractions will also result in less CO flux.

Figure \ref{fig:Fco_comp} compares results for the average CO signal as a function of halo mass produced by the full \citet{Obreschkow09b} model---including both equation \ref{eq:Ob} and their semi-analytic prescriptions for galaxy formation---to ours assuming $Z'=0.1$ and a fiducial LSW angular momentum transport model with $m=0.2$.  Note however, that the CMB has been subtracted in their results, while we plot only $\Fco$ from our model.  The \citet{Obreschkow09a} method predicts a comparable amount of CO emission to our model if we include the full effect of turbulent clumps on the level populations.  Additionally, their signal is equally difficult to observe when compared to the observational sensitivities of JVLA or ALMA.  However, given our very different methodologies, the similarity between our results is probably coincidental.  For example, an even higher-order CO line (such as $J=10$) would likely be very sub-thermal in our model even with the higher median cloud densities produced by turbulent clumps, while the \citet{Obreschkow09b} model would predict much more emission at possibly even detectable levels based on their assumption of thermal equilibrium.  Moreover, the effect of metallicity in our model is quite non-linear, in sharp contrast with their prescription.

For a more physical comparison of our two models, we explore the theoretical underpinnings of equation \ref{eq:Ob} and how it relates to our physically-derived method.  We first remove the extra complication of the CMB-subtracted observation and focus on the emission produced.  Since the amount of CMB-subtraction negligible for those local galaxy from which equation \ref{eq:Ob} was calibrated, then the intrinsic CO emission is well-fit by ignoring the second term of equation \ref{eq:lOb}.  We can further simplify this new model for the production (rather than the observation) of CO emission by assuming optically thick lines at a fixed temperature, $\texc$, which agrees with equation \ref{eq:tauOb} for $J=1$ and $\texc\lesssim30\,\K$.  Finally, since \citet{Obreschkow09b} find that cloud overlap and smooth molecular gas are not important at low redshift and roughly cancel out at high redshift, we set $B'=\epsilon'=1$.  The resulting emission derived from equation \ref{eq:Ob} is
\begin{equation}\label{eq:Ob1}
\tilde{L}_{J} = k\,Z\,M_{\rm H_{2}, tot}\,\frac{J^4}{{\exp}\left(\frac{h_{\rm p}\,\nuco\,J}{\kb\,\texc}\right)-1}.
\end{equation}
By contrast, our method involves summing the luminosity from all clouds as a function of galactocentric radius.  Under the assumption of thermal, optically thick emission with $\texc=\tcmb$ appropriate for equation \ref{eq:Ob1}, this gives 
\begin{equation}\label{eq:L1}
\tilde{L}_{J} = \frac{16\,\pi\,\hp\,\nuco^4}{c^3}\,\left<\frac{\fco}{\fgmc}\,\frac{\sigcl}{\Scl}\right>\,M_{\rm H_{2}, tot}\,\frac{J^4}{{\exp}\left(\frac{h_{\rm p}\,\nuco\,J}{\kb\,\texc}\right)-1},
\end{equation}
where $\left<\fco\,\sigcl/\fgmc\,\Scl\right>$ is the molecular-mass--weighted average of $\fco\,\sigcl/\fgmc\,\Scl$ throughout the galaxy.  Equating equations \ref{eq:Ob1} and \ref{eq:L1} and assuming $Z=0.01\,Z'$ gives a value for $k$ of
\begin{equation}\label{eq:k}
k = 6.15\times10^{-8}\,{\rm W\,kg^{-1}}\,{Z'}^{-1}\,\left<\frac{\fco/\fgmc}{0.5}\,\frac{85\,{\rm \msun\,pc^{-2}}}{\Scl}\,\frac{\sigcl}{1\,{\rm km\,s^{-1}}}\right>.
\end{equation}
Thus, our methodology can reproduce the normalization factor that \citet{Obreschkow09b} determine empirically under similar assumptions.  However, our model also provides results when those assumptions are relaxed and, in particular, allows for a realistic extrapolation to the physical conditions at high redshift.  For metallicities in the predicted range of $Z'=0.03$--0.3, $k$ can be significantly lower---depending on halo mass and disk model---than the value used by \citet{Obreschkow09b} because of the strong metallicity dependence of ${\fco/\fgmc}$.  Note that, in our model, $\sigcl\,\Scl^{-1} \propto \Omega^{-1}$ if $\Sgas>85\,{\rm \msun\,pc^{-2}}$ (combining equations \ref{eq:Mcl} and \ref{eq:sigma_cl}), but we leave them as independently specified in equation \ref{eq:k} for generality.

More recently, \citet{Lagos12} also employed a semi-analytic model of galaxy formation but coupled it to a PDR calculation that considers the dissociation of CO.  While taking a more sophisticated approach than \citet{Obreschkow09b}, the lack of resolution on small scales necessitates assumptions about the ISM gas that may not be applicable to very high redshifts.  For example, these authors use the same determination of the molecular fraction from equation \ref{eq:Rmol-P} as does \citet{Obreschkow09b} and, additionally, fix $A_{V}=8\,{\rm mag}$ and $n_{\rm H}=10^{4}\,{\rm cm^{-3}}$ by hand, independent of metallicity, galactic radius, or galaxy mass, while we have shown that GMC properties vary strongly with these parameters (see Fig. \ref{fig:gmc}).  This treatment implicitly ignores structure within the turbulently-supported clouds themselves, which we have shown has a strong influences on the thermalization of higher-order CO lines (see Fig. \ref{fig:FVco}).  Finally, the many effects of the CMB at very high redshift are also ignored in this study.  Nevertheless, their Table 4 reports $J=6$ fluxes as a function of rest-frame UV magnitude that are comparable to the results in our Figure \ref{fig:Fco_vs_Muv}, though a precise comparison with their average results is difficult to determine from their four selected objects.

%-------------------------------------------------------------------------------------------------------------
%       Conclusions
%-------------------------------------------------------------------------------------------------------------
\section{Conclusions}\label{sec:conclusions}

In \citet{MF12} we improved and tailored a one-zone ISM model, which had already been found to describe lower redshift systems, for use at $z\gtrsim6$.  In the present work, we used this specialized formalism to calculate the properties of GMCs and predict their CO line fluxes in a way that realistically considers the physical conditions expected in Lyman-break galaxies at very high redshift.  Note that we do not consider sub-millimeter galaxies and quasar hosts, which may not be well-described by our models.

Comparing GMC properties for three different fiducial angular momentum transport mechanisms---linear spiral waves (LSWs) with $m=0.2$ and $m=1$ and nonlinear infall with $\beta=0.01$---we found differences in the radial distributions of cloud mass, number, and visual extinction among our models.  In particular, the properties for clouds in the $m=1$ LSW model are more constant with galactocentric radius than for other models because of a higher velocity of centrally-flowing gas.  We then proposed a new method for determining $\fgmc$, taking the star formation rate surface density, the gas surface density, and the star formation efficiency in clouds as fundamental inputs.  Comparing our results with the chemical equilibrium calculation of \citet{Krumholz09b} suggests either that the metallicity of $z\gtrsim6$ galaxies is of order a few percent of solar or that the clouds in these systems are somewhat smoother than at $z=0$.  We then used a simple PDR model of clouds from \citet{Wolfire10} to investigate the effect of metallicity on the region within the cloud capable of maintaining its carbon in the form of CO.

We set the gas temperature to be a simple combination of the local effective temperature of the disk and the CMB temperature at the given redshift.  Using this temperature and taking into account the higher average densities in the central, CO-containing regions of clouds, we employed a modified version of the \citet{KT07} escape probability code to calculate the CO emissivity with no {\it{a priori}} assumptions on the level populations or optical depths of the transitions.  We computed the resulting flux from the entire galaxy by simply counting up the number of clouds at each galactic radius and found that, the results for the $J=6$ transition strongly depend on the presence of turbulent clumps within clouds and their influence on the population of states; the $J=1$, however, is relatively insensitive to this effect.

Comparing our results to the universal model for the CO X-factor predicted by \citet{Narayanan12}, we found very good agreement in regions that those authors simulated when we adjusted the CMB temperature to compensate for their different redshifts of interest.  While our predictions are also comparable to those from the semi-analytic method of \citet{Obreschkow09b}, we suggest that this agreement is largely coincidental given our very methodology.  We would have expected a strong disparity due to our improved treatments of molecular gas fractions, the dependance of CO-dissociation on metallicity, and CO level populations and optical depths.  We showed that our model can reproduce their locally-calibrated normalization constant for CO emission if we adopt a simplified set of assumptions, and that this constant should be a function of galaxy properties in ways not accounted for by their model.

Our final determination of the CMB-subtracted, velocity-integrated CO flux as a function of halo mass and redshift depends on metallicity but is fairly robust to the choice of angular momentum transport model in the galactic disk.  We predict that the CO signal will be very difficult to observe either with existing or currently-planned facilities.  Only the largest galaxies may be detectable with reasonable observation times and only if both $Z'\gtrsim0.1$ and turbulent clumps within clouds affect the thermalization of the $J=6$ transition as strongly as we have assumed.  Blind searches for new populations of high-redshift galaxies will be particularly inefficient given the rarity of these bright objects.  However, targeted studies may prove fruitful if a significant amount of observing time is dedicated to extremely UV-bright candidates.  Because of this, shallow, wide-area observations in the rest-frame UV, such as those by the HIPPIES \citep{HIPPIES10} and BoRG \citep{BoRG11} programs, or high-magnification lensing surveys like CLASH \citep{Zitrin11} or the upcoming {\it{Hubble}} Frontier Fields will be the most useful sources of candidates suitable for followup.  While, our results also imply that the integrated CO luminosity of the entire high-redshift galaxy population is small, we will leave a more thorough treatment of the feasibility of ``intensity mapping" of CO lines from unresolved galaxies \citep{Lidz11, Carilli11, Gong11} to future work.

In a forthcoming paper, we will continue to probe the observational signatures of this model and capitalize on its ability to calculate GMC properties and molecular line emission.  In particular, given the dissociated state of much of the carbon in low metallicity GMCs, we will apply our formalism to the problem of CII emission lines from high-redshift galaxies.  These lines have already been observed out to $z\sim5$ in bright submillimeter galaxies \citep[e.g.][]{Wagg12}.  Predictions for more typical galaxies out to even higher redshifts will allow us to explore the earliest era of galaxy formation in the universe.

%-------------------------------------------------------------------------------------------------------------
\section{Acknowledgements}

We thank Jean Turner, Desika Narayanan, Chris Carilli, Mark Krumholz, and Danail Obreschkow for helpful discussions and suggestions.  This research was partially supported by the David and Lucile Packard Foundation and the Alfred P. Sloan Foundation.

%-------------------------------------------------------------------------------------------------------------
%-------------------------------------------------------------------------------------------------------------
%       Appendix 
%-------------------------------------------------------------------------------------------------------------
%-------------------------------------------------------------------------------------------------------------
\begin{appendix}

%-------------------------------------------------------------------------------------------------------------
\section{Modifications to the Krumholz \& Thompson Code} \label{sec:app:KT}
In this section, we document the changes we made to the publicly-available escape-probability code of \citet{KT07}.
 
\subsection{{\it{J}}=1--0 Optical Depth} \label{sec:app:KT:tau}

The original version of the code takes the optical depth of the $J=1$ transition as an input and determines $\tau_{J}$ for each molecular line as a fraction of $\tau_{J=1}$.  As such, it only does part of the calculation for $\tau_{J}$ before using $\tau_{J=1}$ as a normalization constant.  However, because the $J=1$ transition is not necessarily in thermal equilibrium, its optical depth is difficult to calculate.  We alter the code to compute $\tau_{J}$ without the normalization constant, which requires introducing the radius of the CO-containing region of the cloud as an additional input.

\subsection{CMB} \label{sec:app:KT:cmb}
The original version of the code ignores the ability of the CMB to radiatively excite or de-excite levels of CO in a similar way to radiation from the CO itself.  The absorption rate of CMB photons per particle is 
\begin{eqnarray}\label{eq:CMBabs}
\Gamma_{\rm CMB}&=&\beta_{J}\,B_{J-1{\rm ,} J}\,B(J\,\nuco, \tcmb)\nonumber\\
&=&\beta_{J}\,\frac{g_J}{g_{J-1}}\,\frac{A_{J{\rm ,} J-1}}{{\rm e}^{h_{\rm P}\,J\,\nuco/\kb\,\tcmb}-1},
\end{eqnarray}
where $B_{J-1{\rm ,} J}$ is the Einstein--B coefficient for absorption and $h_{\rm P}$ and $\kb$ are the Planck and Boltzmann constants, respectively.  In an optically thin region (i.e., $\beta_J=1$), the CMB absorption rate is more significant for lower-order transitions.  However, because an increase in a particular level necessitates a decrease elsewhere in the distribution, all levels will be somewhat affected.  To modify the code, we note that the excitation rate per particle due to emission from CO is $\beta_{J}\,A_{J{\rm ,} J-1}$, where the factor of $\beta_{J}$ accounts for line photons that escape the cloud without contributing to the excitation.  Therefore, we simply multiply $\beta_{J}\,A_{J{\rm ,} J-1}$, where it appears in the code, by an additional factor of $(1+[g_J/g_{J-1}]/[{\rm e}^{h_{\rm P}\,J\,\nuco/\kb\,\tcmb}-1])$.  Ultimately, this modification reduces the luminosity in the lines only by, at most, a factor of a few.

\end{appendix}
%-------------------------------------------------------------------------------------------------------------
\bibliography{co_highz_rev1.bbl}

%-------------------------------------------------------------------------------------------------------------
\end{document}